\documentclass[traditabstract]{aa}
\usepackage{psfig}
\usepackage{natbib}
\usepackage{lscape}
\usepackage{amsmath}
\usepackage{wasysym}
\usepackage{graphics}
\usepackage{txfonts}
\usepackage{color}
\usepackage{times}
\usepackage{color,colortbl}

\usepackage{hyperref}
\hypersetup{
    bookmarks=true,         
    unicode=false,          
    pdftoolbar=true,        
    pdfmenubar=true,        
    pdffitwindow=true,      
    pdftitle={Fitting Analysis using Differential Evolution Optimization},    
    pdfauthor={Gomes \& Papaderos},     
    pdfsubject={FADO},      
    pdfnewwindow=true,      
    pdfkeywords={keywords}, 
    colorlinks=true,        
    linkcolor=blue,         
    citecolor=blue,         
    filecolor=blue,         
    urlcolor=blue           
}

\newfont{\vssn}{cmss10 scaled 1050}
\newfont{\vsss}{cmss10 scaled 450}
\newfont{\nsf}{cmssdc10 scaled 1000}
\newfont{\vs}{cmssdc10 scaled 700}
\newfont{\lvss}{cmssdc10 scaled 900}
\newfont{\lx}{cmssdc10 scaled 760}
\newfont{\llx}{cmssdc10 scaled 1000}
\newfont{\nlx}{cmssdc10 scaled 900}
\def\rem{\nlx}

\def\eqan{\begin{equation}}
\def\eqen{\end{equation}}

\def\h1{\ion{H}{i}}
\def\h2{\ion{H}{ii}}

\newcommand{\zsun}{$Z_\odot$}

\newcommand{\oh}{12+log(O/H)}
\def\ha{H$\alpha$}

\def\hb{H$\beta$}

\def\hd{H$\delta$}

\def\o5007{[O {\sc iii}] $\lambda$5007}

\def\P25{{\sl R}$_{\rm SF}$}
\def\E25{{\sl R}$_{\rm host}$}
\def\ige{\nsf ne\rm}

\newcommand{\PutLabel}[3]{\put(#1,#2){#3}}
\def\mstar{${\cal M}_{\star}$}

\def\tmass{$\langle \log t_\star \rangle_{{\cal M}}$}
\def\tlum{$\langle \log t_\star \rangle_{{\cal L}}$}
\def\Zmass{$\langle Z_\star \rangle_{{\cal M}}$}
\def\Zlum{$\langle Z_\star \rangle_{{\cal L}}$}

\def\BaseS{Base ${\cal S}$}
\def\BaseF{Base ${\cal F}$}

\def\FADO{{\sc \textbf{fado}}}
\def\starlight{{\sc Starlight}}

\def\FCmode{{\nsf FCmode}}
\def\NCmode{{\nsf NCmode}}
\def\STmode{{\nsf STmode}}

\def\D4000{$D_{4000}$}

\def\zstar{$Z_{\star}$}

\def\ige{\nlx{ne}\rm}
\def\ess{\nlx{ess}\rm}
\def\pss{\nlx{pss}\rm}

\def\uflux{erg\ s$^{-1}$\ cm$^{-2}$}
\newcounter{qub}
\usepackage{colortbl}
\definecolor{orange1}{rgb}{1,0.5,0}
\definecolor{cyan2}{rgb}{0.6,0.18,0.49}
\definecolor{myblue1}{rgb}{0.0,0.604,0.831} 
\definecolor{myblue2}{rgb}{0.0,0.49,0.6745}
\definecolor{myblue3}{rgb}{0.0156,0.4078,0.9921}
\definecolor{myred1}{rgb}{0.529,0.019,0.017}

\newcommand{\Starlight}{{\sc Starlight}}

\begin{document}

\title{Fitting Analysis using Differential Evolution Optimization ({\FADO}):}

\subtitle{Spectral population synthesis through genetic optimization under
  self-consistency boundary conditions \thanks{The distribution package of the
    FADO v.1 tool contains the binary and its auxiliary files. FADO v.1 is
    available in electronic form at the CDS via anonymous ftp to
    \href{cdsarc.u-strasbg.fr}{cdsarc.u-strasbg.fr} (130.79.128.5) or via
    \href{http://cdsweb.u-strasbg.fr/cgi-bin/qcat?J/A+A/}{http://cdsweb.u-strasbg.fr/cgi-bin/qcat?J/A+A/}}}

\titlerunning{Fitting Analysis using Differential evolution Optimization (\FADO)} 

\author{J.M. Gomes \and P. Papaderos}

\institute{Instituto de Astrof\'{i}sica e Ci\^{e}ncias do Espa\c{c}o,
  Universidade do Porto, CAUP, Rua das Estrelas, PT4150-762 Porto, Portugal}
\date{Received ?? ; Accepted ??}

\abstract{The goal of population spectral synthesis (\pss; also referred to as
  inverse, semi-empirical evolutionary- or fossil record approach) is to
  decipher from the spectrum of a galaxy the mass, age and metallicity of its
  constituent stellar populations.  This technique, which is the reverse of
  but complementary to evolutionary synthesis, has been established as
  fundamental tool in extragalactic research. It has been extensively applied
  to large spectroscopic data sets, notably the SDSS, leading to important
  insights into the galaxy assembly history. However, despite significant
  improvements over the past decade, all current \pss\ codes suffer from two
  major deficiencies that inhibit us from gaining sharp insights into the
  star-formation history (SFH) of galaxies and potentially introduce
  substantial biases in studies of their physical properties (e.g., stellar
  mass, mass-weighted stellar age and specific star formation rate).  These
  are {\rem i)} the neglect of nebular emission in spectral fits,
  consequently, {\rem ii)} the lack of a mechanism that ensures consistency
  between the best-fitting SFH and the observed nebular emission
  characteristics of a star-forming (SF) galaxy (e.g., hydrogen Balmer-line
  luminosities and equivalent widths-EWs, shape of the continuum in the region
  around the Balmer and Paschen jump).  In this article, we present
  \FADO\ ({\rem Fitting Analysis using Differential evolution Optimization})
  -- a conceptually novel, publicly available \pss\ tool with the distinctive
  capability of permitting identification of the SFH that reproduces the
  observed nebular characteristics of a SF galaxy.  This so-far unique
  self-consistency concept allows us to significantly alleviate degeneracies
  in current spectral synthesis, thereby opening a new avenue to the
  exploration of the assembly history of galaxies.  The innovative character
  of \FADO\ is further augmented by its mathematical foundation: \FADO\ is the
  first \pss\ code employing genetic differential evolution
  optimization. This, in conjunction with various other currently unique
  elements in its mathematical concept and numerical realization (e.g.,
  mid-analysis optimization of the spectral library using artificial
  intelligence, test for convergence through a procedure inspired by Markov
  chain Monte Carlo techniques, quasi-parallelization embedded within a
  modular architecture) results in key improvements with respect to
  computational efficiency and uniqueness of the best-fitting
  SFHs. Furthermore, \FADO\ incorporates within a single code the entire chain
  of pre-processing, modeling, post-processing, storage and graphical
  representation of the relevant output from \pss, including emission-line
  measurements and estimates of uncertainties for all primary and secondary
  products from spectral synthesis (e.g., mass contributions of individual
  stellar populations, mass- and luminosity-weighted stellar ages and
  metallicities).  This integrated concept greatly simplifies and accelerates
  a lengthy sequence of individual time-consuming steps that are generally
  involved in \pss\ modeling, further enhancing the overall efficiency of the
  code and inviting to its automated application to large spectroscopic data
  sets.}


\maketitle \markboth {Gomes \& Papaderos}{\FADO: a new avenue to the
  exploration of galaxy formation and evolution}

\section{Introduction \label{intro}}
Understanding the formation and evolution of galaxies is undoubtedly one of
the greatest challenges of modern astronomy. Excluding the few systems in our
close vicinity that can be resolved into individual stars and studied through
color-magnitude diagrams, the star-formation history (SFH) of galaxies can
only be inferred from spectra, each typically containing the
luminosity-weighted output from millions of stars. Deciphering from a spectrum
the SFH and chemical enrichment history (CEH) of a galaxy is the objective of
spectral synthesis - one of the most computationally demanding yet fundamental
tools of extragalactic research.  Spectral synthesis comes essentially in two
reverse yet complementary techniques -- evolutionary and population synthesis
(also referred to as inverse, semi-empirical evolutionary- or fossil record
approach), inspired, respectively, by the seminal works by
\citet{StruckMarcellTinsley1967} and \citet{Faber72}.

The goal of evolutionary spectral synthesis (\ess) is to compute the time
evolution of the spectral energy distribution (SED) of a galaxy on the basis
of an assumed SFH and CEH. Comparison of synthetic with observed SEDs permits
us to place constraints on the galaxy assembly history (for example,
disentangle a quasi-monolithic from a continuous formation process).  The fact
that the SFH is, by definition, an input quantity to \ess\ facilitates
straight-forward predictions of the time evolution of various observables of
interest (e.g., colors, hydrogen Balmer-line equivalent widths) and eases a
detailed treatment of relevant processes contributing to a galaxy SED (e.g.,
reprocessing of the ionizing photon output from OB stars into nebular emission
for various assumptions on the chemistry, geometry and physical conditions of
the gas component).  Indeed, starting from the pioneering work by
\cite{StruckMarcellTinsley1967}, \ess\ underwent several important refinements
that permitted nearly panchromatic predictions of the spectrophotometric
evolution of galaxies, notably the chemically consistent treatment of gas and
stars, and the inclusion of nebular and dust emission in several codes
\citep[e.g.,][among others; see \textcolor{blue}{Tinsley 1980} and, recently,
  \textcolor{blue}{Conroy 2013} for reviews on the
  subject]{Leitherer1995,Krueger1995,FR97,GRASIL,Leitherer99,FR99,Kewley2001,CL01,Zackrisson2001,AF03,Anders04-NGC1569,Groves2008,Zackrisson2008,Kotulla09-GALEV,SchaererBarros09,Verhamme2012,Molla2009,MM10-POPSTAR}
and facilitating different SED fitting tools that include stellar \& nebular
emission
\citep[e.g.,][]{Burgarela2005-CIGALE,Serra2011-CIGALE,Acq11,Pacifici12,Amorin2012,HH14,CC16-Beagle,Leja16-Prospector}.
Such \ess\ models and the fitting tools built upon them have been extensively
applied to the interpretation of galaxy observations
\citep[e.g.,][]{Guseva2001,Moy2001,Guseva2007,Cairos2009-Mrk409,Moustakas2010,Izotov2011-GP,Amorin2012,Schaerer2013,Brinchmann2013,Maseda2014,Pacifici15,RM2015,vdW2016,
  Izotov16-LyC}.

In the past decade, \ess\ saw impressive advances with regard to the
sophistication and complexity of input SFHs\&CEHs.  Following the early era of
$\chi^2$-based searches among a limited set of SEDs corresponding to a few
simple SFH parametrizations (e.g., continuous star formation at a constant
star-formation rate-SFR or exponentially declining SFR with an e-folding
timescale between 0.1 and 10 Gyr), several modern variants of \ess\ exploit
Bayesian or Markov chain Monte Carlo techniques to search within a set of mock
SEDs for the one giving the best match to observations, while additionally
providing the posterior probability distribution of various non-linearly
coupled evolutionary ingredients.  For example, \cite{Pacifici12} employ a
Bayesian approach to a dense grid of pre-computed SEDs, whereas in the models
by both \cite{CC16-Beagle} and \cite{Leja16-Prospector} synthetic SEDs are
computed on the fly.  Approaches taken in computing SED grids include,
for example, Monte Carlo realizations of parametric SFHs with random
bursts optionally added
\citep[e.g.,][]{Kauffmann2003a,Kauffmann2003b,Kauffmann2003c,Tremonti2004,Brinchmann2004,daCunha08_MAGPHYS},
purely non-parametric SFHs (and CEHs) adopted from post-treatment of
cosmological simulations \citep[e.g.,][see also \textcolor{blue}{De Lucia \&
    Blaizot 2007}]{Pacifici12,Pacifici16}, or combinations thereof in various
spectral fitting tools geared toward multi-band photometry and
spectroscopy
\citep[e.g.,][]{Acq11,Pacifici12,HH14,CC16-Beagle,Leja16-Prospector}.  The
trial SEDs used by many of these advanced spectral fitting codes
\citep[e.g.,][]{Brinchmann2004,Acq11,Pacifici12,CC16-Beagle} consistently take
into account the nebular emission expected from the Lyman continuum (LyC)
photon rate from OB stars, reaching in some cases \citep[e.g.,][]{CC16-Beagle}
a significant degree of sophistication in the treatment of nebular physics
(with, e.g., gas ionization parameter, metallicity and dust-to-metal mass
ratio being adjustable input parameters) through coupling of \ess\ with
refined photoionization codes \citep[e.g., CLOUDY;][]{Ferland2013}.  Indeed,
the relatively simple "forward" physical concept of \ess\ -- in essence,
SFH-weighted convolution of simple stellar population (SSP) spectra -- greatly
simplifies inclusion in synthetic SEDs of the post-processed stellar radiation
by gas and dust, at the price of ad hoc or semi-empirically founded
assumptions on the SFH and CEH of galaxies.  Clearly, an important virtue of
such coupled \ess+photoionization codes is the implicit consistency between
the stellar LyC photon output and the predicted Balmer-line luminosities,
given that the former is used for the computation of the latter.  It should be
born in mind, however, that consistency between observed and predicted nebular
emission in this case does not imply that the full set of evolutionary and
physical characteristics of the stellar component of the best-fitting mock SED
(i.e., age and metallicity) are inferred self-consistently, taking into
account the observed nebular characteristics, within a multi-dimensional
topology of non-linearly coupled parameters.

Summarizing, the SFH -- regardless of whether it is entered in a parametric or
non-parametric form -- remains the main input assumption in current
\ess\ models and SED-fitting tools built upon them.  The same applies to the
CEH: for instance, a best-fitting SED with fixed stellar metallicity $Z$
encapsulates in itself a strong assumption on the CEH of a galaxy (no chemical
evolution of stars and gas, in contrast to the chemically consistent
evolutionary approach).  The benefits from consistent inclusion of nebular
emission in chemically inconsistent SED templates might be a subject of
debate, given the strong dependence of the LyC photon rate on $Z$.  For
instance, the \ha\ luminosity expected for a Salpeter initial mass function
(IMF) increases from 2$\cdot$\zsun\ to \zsun/20 by a factor of approximately
four \citep[see e.g.,][and references therein]{WF01}, which is certainly of
relevance both to SED fitting and SFR determinations of chemically evolving
galaxies across cosmic time.

Conversely, the goal of population spectral synthesis \citep[\pss, see][for a
  review]{Walcher2011} is to decompose the observed spectrum of a galaxy into
its elementary and no further divisible constituents. These are spectra of
individual stars or SSPs (spectral snapshots of instantaneously formed stellar
populations being fully characterized by their age $t$, chemical composition
and initial mass function) that a \pss\ code selects from a spectral base
library.  The best-fitting solution is an array holding the mass fractions
$\mu_j$ of the SSP "building blocks" picked up from the library and is
referred to as population vector\footnote{In the following, we will use the
  term "population vector" in a more generic sense, in order to denote the
  full solution of a \pss\ code, that is, the best-fitting
  velocity-dispersion-convolved combination of spectral ingredients (SSPs,
  nebular emission) and their respective extinction (see Sect.~\ref{FADO:DEO}
  for details).} (PV).  The array $\mu_j$(SSP($t$,$Z$)) obviously yields a
discretized approximation to the SFH and CEH of a galaxy.  From the seminal
work of \citet{Faber72} and the remarks above it is apparent that \pss\ is a
spectral decomposition (aposynthesis) technique, and inverse to
\ess\ (synthesis).  Quite importantly, the only condition in this
"mathematical optimization exercise" is that the $\mu_j$ elements composing
the best-fitting PV are positive and finite. In particular, \pss\ is by
definition irreconcilable with any prior or implicit assumption on the SFH and
CEH: even a simple functional relation binding two SSPs of different age (or
$Z$) into one single "SFH block" would be equivalent to an implicit assumption
on the SFH (or CEH), and would thus violate the principles of \pss.  The
"inverse" approach of \pss\ obviously also entails a different mathematical
concept than that of "forward computing" \ess\ codes, and makes inclusion and,
the more so, consistent treatment, of nebular emission a challenging task
within the mathematical/numerical domain of multi-objective optimization
(cf. Sect.~\ref{FADO:DEO}).  Finally, given the considerable confusion in the
nomenclature (with several cases in the literature citing the term \ess\ and
\pss\ interchangeably, or associating \pss\ with \ess\ codes using
non-parametric input SFHs/CEHs) it is useful to bear in mind that the term
"parametric" or "non-parametric" is strictly speaking inapplicable to \pss,
since the SFH and CEH are not input parameters.

The past decade saw the development of different \pss\ codes, each conceived
with a different priority in mind, such as MOPED \citep{Heavens00}, pPXF
\citep{CE04}, \starlight\ \citep{CidFernandes2005,Lopez2016} and STECKMAP
\citep{Ocv06}.  UlySS \citep{Koleva09} might be regarded as a special case,
since its mathematical concept complies with \pss, however, assumptions are
imposed on the best-fitting SFH and CEH.  For instance, the goal of MOPED is
to collapse a galaxy spectrum into its most robust and characteristic
spectroscopic features, whereas ULySS and STECKMAP are primarily intended to
spectral fitting and kinematical analyses. \starlight, on the other hand, may
arguably be regarded as the most intensively applied \pss\ code for the
exploration of the SFH of galaxies on the basis of large spectroscopic data
sets \citep[e.g., the SDSS,][]{York2000}.

However, despite conceptual and numerical improvements, all currently
available \pss\ codes suffer from two important deficiencies that limit their
potential for gaining sharp insights into the galaxy assembly history.  The
first one stems from the notorious SFH -- extinction -- metallicity degeneracy
\citep[e.g.,][]{OConnell1996,Pelat1997,Pelat1998} and the second one, as we
argue next, from the neglect of nebular emission in spectral fits.  A direct
consequence from the second is obviously the lack of a mechanism that ensures
consistency between the best-fitting PV and the observed nebular emission
(\ige) characteristics in a galaxy, as for example the luminosities and
equivalent widths (EWs) of Balmer recombination lines, and the SED
characteristics in the spectral region around the Balmer and Paschen jump
(3646\AA\ and 8207\AA, respectively).

{\rem a) Degeneracy of spectral fits:} A well known yet far from overcome
problem in spectral synthesis (both to \pss\ and to \ess) is that an
irreproachable fit is in itself no proof neither for the uniqueness nor
astrophysical soundness of the best-fitting model: in fact, substantially
different PVs, composed of SSPs with different age and \zstar, and subjected
to different amounts of intrinsic extinction can result in almost
indistinguishable SEDs \cite[see, e.g.,][hereafter G01, for a related
  discussion in the case of grid-based SED-fitting \ess\ tools]{Guseva2001}.

This is specially relevant to studies of star-forming galaxies, where the
imprints of, for example, a small (2--5\%) mass fraction of instantaneously
formed young ionizing stars on the optical continuum can readily be reproduced
by a slightly older episode of prolonged star formation without any
appreciable young ionizing stellar component (cf. G01).  Such degeneracies in
the PV may translate into substantial (20-50\%) uncertainties in the burst
parameter\footnote{The mass fraction of stars formed in a starburst, as
  compared to the stellar mass ever formed
  \citep[e.g.,][]{Leitherer1995,Krueger1995}.} {\sl b}, propagating then into
other quantities of interest, such as the specific SFR (sSFR).

More generally, as shown by an examination of this issue by G01, the
uniqueness of spectral fits for star-forming galaxies can hardly be
established from $\chi^2$-minimization techniques to the optical continuum
alone, that is without additional constraints (e.g., EWs of Balmer absorption
and emission lines).  The degeneracy problem is probably further
aggravated by various technical specifics of the fitting procedure (e.g.,
spectral range considered, quality of the kinematical fitting of stellar
absorption features, construction of the SSP library; cf Appendix) the
combined effect of which on \pss\ modeling has not been conclusively
investigated as yet, even though pilot attempts in this direction exist
\citep[see, e.g.,][\textcolor{blue}{Cardoso, Gomes \& Papaderos, in prep.,
    hereafter
    CGP17}]{Gomes2005,Gomes2009,Chen2010,Richards2009,CidFernandes2014,Magris2015}.

{\rem b) Neglect of nebular emission:} A fundamental conceptual shortcoming of
all state-of-the-art \pss\ codes is the neglect of \ige.  First, \ige, as
natural consequence of star formation is inseparable from the galaxy assembly
process and an indispensable ingredient of any physically meaningful spectral
model \citep[e.g.,][]{Grewing1968,Huchra77,Leitherer1995,Krueger1995}.  In
particular, inclusion of \ige\ is crucial to the \pss\ modeling of
intermediate-to-high $z$ galaxies, where intense SF activity is virtually
omnipresent. According to our current knowledge, these systems are building up
their stellar component through starbursts or prolonged episodes of strongly
elevated sSFR, translating into short (a few $10^8$~yr) stellar mass (\mstar)
doubling times.  The ionizing output from massive young stars, forming at
prodigious rates during such dominant phases of galaxy buildup is plausibly
expected to excite strong \ige\ on kpc scales and boost EWs of strong nebular
emission lines (e.g., [O{\sc iii}]$_{5007}$ and \ha) to values exceeding
$10^3$ \AA. For instance, \citet{Krueger1995} show that in a strong ($b=0.1$)
SF episode \ige\ contributes between $\sim$30\% and $\sim$70\% of the total
optical and near-infrared (NIR) emission.  Blue compact dwarf (BCD) and
several \h2\ galaxies
\citep[e.g.,][]{LooseThuan1986,Salzer89,Terlevich1991,P96a,P96b,LM01a,BergvallOstlin2002,GildePaz03-BCDs,SanchezAlmeida2008},
and their higher-$z$ analogs \citep[e.g., compact narrow-emission line
  galaxies and green peas;][among
  others]{Koo1994,Guzman1998,Puech2006-LCBs,Cardamone2009-GP,Amorin2010-GP,Izotov2011-GP,Atek2011,vdW2011,Amorin2012,JaskotOey13-GP,Amorin2015-emlin}
offer examples of the substantial contamination of starburst galaxy SEDs by
\ige. The most extreme cases in the local universe are arguably extremely
metal-poor (\oh~$\leq$~7.6) BCDs where intense and galaxy-wide SF activity in
conjunction with the low surface density of the underlying stellar background
give rise to \ha\ and [O{\sc iii}]$_{5007}$ EWs as high as $\sim$2000
\AA\ \citep[see,
  e.g.,][]{Izotov97-SBS0335,Izotov97-IZw18-WR,Papaderos1998-SBS0335,Izotov2001-SBS0335,Fricke01-Tol1214,Papaderos2002-IZw18,I04-T1214-T65,Izotov09-SBS0335}.

As discussed in \citet{PapaderosOstlin2012} the neglect of \ige\ in spectral
modeling studies of high-sSFR galaxies near and far could introduce
substantial (0.4\dots1~mag) biases in commonly studied galaxy scaling
relations that involve total magnitudes (e.g., the Tully-Fisher relation, or
relations between luminosity and metallicity, diameter, mean surface
brightness and velocity dispersion). Nebular emission also affects
\mstar\ determinations via theoretical M/L ratios or SED fitting: the usual
procedure of flagging strong emission lines (e.g., [O{\sc iii}]$_{4959,5007}$,
\ha, \hb) prior to \pss\ modeling is an inadequate remedy to the problem,
since it does not decontaminate SEDs from the reddish nebular continuum
emission, which in galaxies with strong ({\sl b}=0.1) starburst activity can
exceed 20\% (50\%) of the monochromatic luminosity of the SED continuum in the
$I$ ($K$) band \citep{Krueger1995}.

Specially relevant to \pss\ modeling of starburst galaxies is also the fact
that the nebular continuum has a flatter SED than the young ionizing stellar
component, it thus becomes progressively important with increasing wavelength
$\lambda$ (i.e., it cannot be treated as an achromatic additive offset
to the stellar SED). \citet{Izotov2011-GP} pointed out that this fact may
cause purely stellar models to invoke a much too high (up to a factor of 4)
contribution from old stars, leading to a systematic overestimation of
\mstar\ (correspondingly to an underestimation of the sSFR).  A secondary
concern is that dilution of stellar absorption features by the \ige\ continuum
could bias stellar velocity dispersion measurements with state-of-the-art
(i.e., purely stellar) \pss\ codes.  Even though the considerations
above primarily apply to high-sSFR (starburst) galaxies, the inclusion of
nebular continuum is important for an accurate \pss\ modeling of average
late-type galaxies that form stars at a relatively calm pace \citep[see,
  e.g.,][for statistical studies of the SFR and sSFR in the local
  universe]{Brinchmann2004,Lee2007}.

Evidently, no \pss\ code, regardless of its mathematical sophistication can
compensate for the lack of important SED ingredients (nebular continuum),
conversely, SED fitting with an incomplete set of physical ingredients
unavoidably bears the risk of systematic biases in the obtained SFH, CEH and
\mstar.  In the light of the cautionary remarks above, it is worth bearing in
mind that the extensive body of work employing \pss\ modeling of large
extragalactic data sets \citep[e.g., SDSS, CALIFA, MaNGA,
  SAMI;][respectively]{York2000,Sanchez2012-CALIFA,Bundy2014-MANGA,Croom2012-SAMI}
in the last years
\citep[e.g.,][]{Panter2003,CidFernandes2005,Panter2007,Tojeiro2007,Asari2007,CidFernandes2007,Zhong2010-LSB,Zhao2011-BCDs,LaraLopez2010,Torres-Papaqui2012,SanchezAlmeida2012,Yoachim2012,Perez2013-CALIFA,SanchezJanssen2013,Martins2013,GonzalezDelgado2014a-CALIFA,GonzalezDelgado2014b-CALIFA,SB14-CALIFA,Belfiore2015-MANGA,Lopez2016,Schaefer2016-SAMI}
uses purely stellar SEDs, it therefore relies on the assumption that nebular
continuum emission is invariably negligible in SF galaxies.

In this article, we present \FADO\ (Fitting Analysis using Differential
Evolution Optimization), a conceptually novel publicly available\footnote{A
  thoroughly documented version of the code is available at
  \href{http://www.spectralsynthesis.org}{\tt
    http://www.spectralsynthesis.org}.} \pss\ tool that incorporates the
physical ingredients and mechanisms that ensure consistency of the
best-fitting PV with the observed nebular emission characteristics in a
galaxy. This article is organized as follows: in Sect.~\ref{FADO1} we provide
an overview of the novel concept and key distinctive properties of \FADO\ and
its modular structure, and Sect.~\ref{FADO:DEO} presents a concise description
of its mathematical foundation, in particular of some important advantages of
differential evolution optimization (DEO) algorithms, with further
improvements made in the framework of this project, with respect to standard
optimization approaches for tackling multi-parameter spectral modeling
problems.  Section~\ref{FADO:PF} describes the underlying physical concept of
\FADO\ with special focus on studies of star-forming galaxies, followed by a
presentation of its different spectral fitting modes (Sect.~\ref{FADO:EmLin})
and the computation and storage of the model output and by-products
(Sect.~\ref{FADO:Output}). Section~\ref{FADO:three} presents illustrative
examples of the application of \FADO\ on studies of galaxy spectra with
various characteristics (star-forming, composite, LINER/retired and
passive/lineless) and a summary is given in Sect.~\ref{Summary}.

\section{\FADO: a novel approach to the exploration of the SFH of galaxies \label{FADO1}}
\subsection{\FADO\ in a nutshell \label{FADO_in_a_nutshell}}

\FADO\ is a conceptually novel \pss\ code that, inter alia, permits
identification of the PV that best reproduces the main nebular characteristics
of a star-forming galaxy, more specifically the observed Balmer line
luminosities and EWs, as well as the shape of the continuum at the region
around the Balmer and Paschen jump.  This so far unique self-consistency
concept allows us to significantly alleviate biases in the determination of
physical and evolutionary properties (e.g., \mstar, light- and mass-weighted
stellar age, SFH and sSFR) of SF galaxies with state-of-the-art \pss\ models
(see \textcolor{blue}{CGP17} for a quantitative assessment of this issue and
Appendix for illustrative examples) thereby opening a new window to the
exploration of the assembly history galaxies.  The innovative character of
\FADO\ is further augmented by its mathematical foundation and modular
architecture: \FADO\ is the first \pss\ code employing genetic DEO
(Sect.~\ref{FADO:DEO}). This results in key improvements with respect to the
uniqueness of spectral fits and the overall efficiency of the convergence
schemes integrated in the code. Moreover, \FADO\ is optimized toward
multi-core CPU architectures and allows for handling of up to 2000 base
elements (i.e., SSPs or individual stellar spectra) with up to 24000
wavelength elements each, which is an important advantage toward future work
with higher-resolution SSP libraries.  Important is also that \FADO\ allows
for the determination and storage of emission-line fluxes and EWs, and,
through a build-in routine based on PGplot\footnote{See
  \href{http://www.astro.caltech.edu/~tjp/pgplot}{http://www.astro.caltech.edu/$\sim$tjp/pgplot}
  for details.} permits visualization of the modeling results.\\

Some distinctive conceptual (astrophysical \& mathematical) advantages of
\FADO\ over currently available \pss\ codes are:

\begin{itemize}
\item[1)] On-the-fly computation and inclusion of the nebular continuum
  contribution to the best-fitting SED and identification of the PV (ages,
  metallicities and mass fractions of individual SSPs, intrinsic extinction)
  that reproduces best the nebular characteristics of a galaxy.

\item[2)] Automatic characterization of the input spectrum for the sake of
  optimization of the SSP library and spectral fitting strategy using
  Artificial Intelligence (AI).

\item[3)] Computation and storage of uncertainties for both the full
  information encoded in the best-fitting PV and secondary products (e.g.,
  mass-weighted stellar age and metallicity, emission-line fluxes and EWs),
  and automated spectroscopic classification based on diagnostic emission-line
  ratios after correction for underlying stellar absorption.

\item[4)] On-the-fly determination of the electron temperature and density of
  the ionized gas (whenever possible and meaningful).

\item[5)] Independent determination of the intrinsic extinction in the stellar
  and nebular component.

\item[6)] Stability, quick convergence and high computational efficiency
  thanks to a refined numerical realization of genetic DEO and several Fortran
  2008 compiler features (e.g., internal quasi-parallelization).

\end{itemize}

\subsection{Main modules of \FADO\ \label{FADO1-main}}

The main components and innovations of \FADO\ will be briefly described below
and can be followed in Fig.~\ref{fig:FADOscheme}, which is divided into three
main parts: $\alpha$) Pre-processing of spectral data, $\psi$) Spectral
synthesis through genetic DEO, and $\omega$) Computation and storage of the
model output. We will start with the functionality of the modules involved in
the data pre-processing prior to SED fitting:

\begin{sidewaysfigure*}
\begin{picture}(18.0,17.0)
\put(0.0,0.0){\includegraphics[trim={1cm -3cm 1cm 1cm},clip,scale=1.0,angle=0, width=25cm]{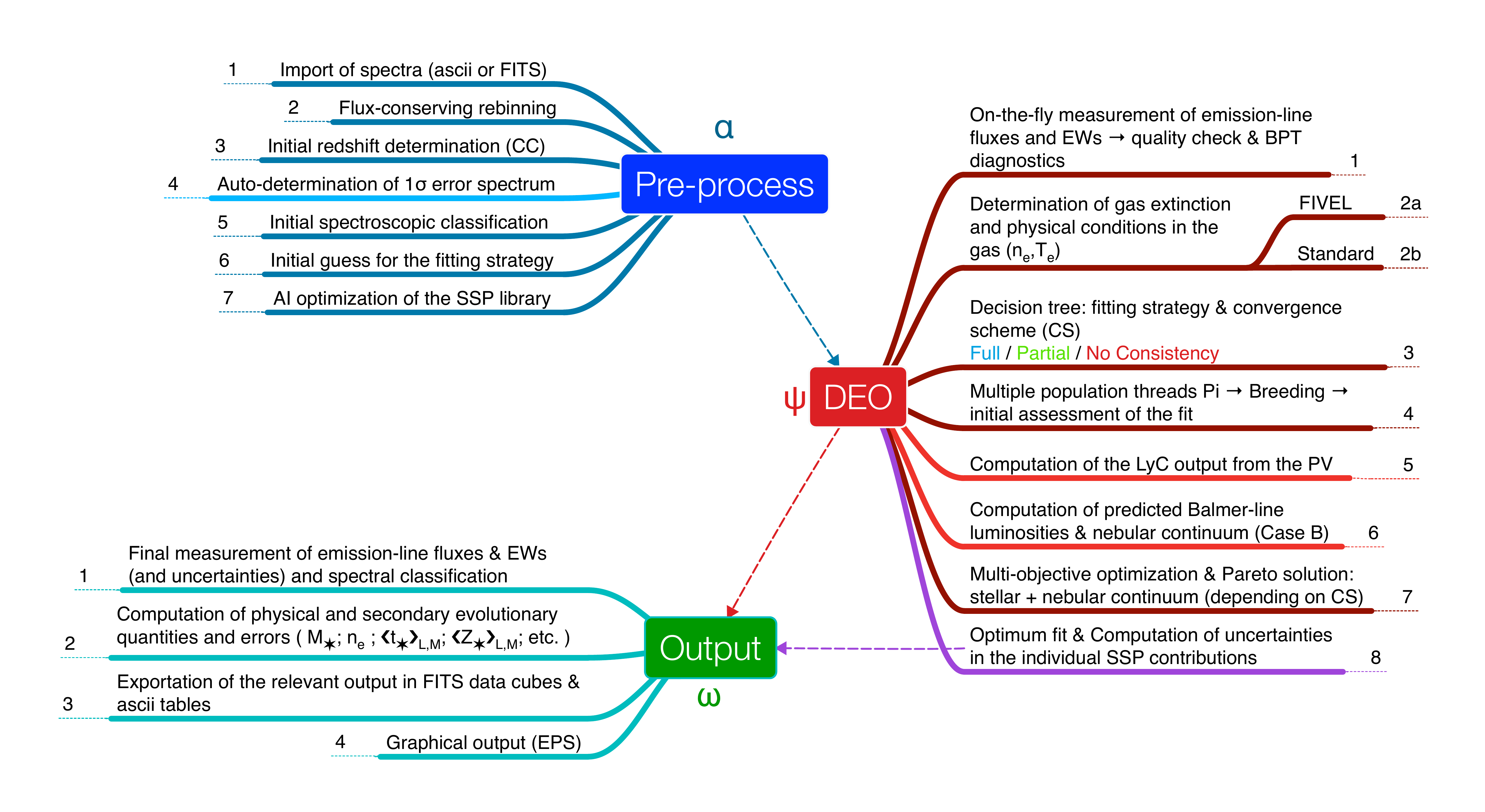}}
\end{picture}
\caption[]{Schematic view of \FADO\ and strategies for reaching convergence}
\label{fig:FADOscheme}
\end{sidewaysfigure*}

\begin{itemize}
\item[$\alpha$1)] {\rem{Import of spectra}}: Ability to handle large (up to
  24k wavelength elements) spectra in both ascii and FITS format.

\item[$\alpha$2)] {\rem{Flux-conserving rebinning}}: When necessary, automatic
  application of flux-conserving rebinning of non-evenly sampled spectroscopic
  data.
   
\item[$\alpha$3)] {\rem{Initial redshift determination}}: Automatic
  determination of the redshift of the source via a cross-correlation
  technique involving stellar absorption features and emission lines.
 
\item[$\alpha$4)] {\rem{Auto-determination of the error spectrum}}:
  Optionally, computation of 1$\sigma$ errors for the input spectrum using
  Tukey's biweight\footnote{Also known as the bisquare function:
\begin{equation*}\beta = \begin{cases}\begin{split} x \left( 1-\frac{x^2}{c^2} \right)^2, \textrm{if |x| < |c|} \\ 0, \textrm{if |x| > |c|}\end{split}\end{cases}\end{equation*}
where $x \equiv [y_i - y(x_i)]/\sigma_i$, with $y_i$ and $y(x_i)$ denoting,
respectively, a measured and predicted value, and $\sigma_i$ a weight factor.
The optimal value for the constant $c$ is 2.1 \citep{Press1992-NR}.} technique
  in robust regression statistics for eliminating outliers \citep[for more
    details see the classical books on robust
    statistics][]{Huber1981,Hampel1986}.

\item[$\alpha$5)] {\rem{Preliminary spectroscopic classification}}: Upon
  initial measurement of emission-line fluxes and EWs on the input spectrum,
  tentative spectral classification according to BPT diagnostics.

\item[$\alpha$6)] {\rem{Initial guess for the fitting strategy}}: Automatic
  adaptation of the fitting strategy (cf. Sect.~\ref{FADO:EmLin}), depending
  on the spectral classification (cf. $\alpha$5) and signal-to-noise (S/N) of
  the input spectrum. As an example, starting from the full-consistency mode
  ({\rem FCmode}), which is set by default, \FADO\ may auto-switch to the
  partial-consistency mode (nebular-continuum mode; {\rem NCmode}) in case of
  a LINER/Seyfert spectrum, or when Balmer emission line fluxes and EWs do not
  fulfill certain quality criteria.

\item[$\alpha$7)] {\rem{Optimization of the SSP library through Artificial
    Intelligence}}: For spectral libraries exceeding 800 SSPs, \FADO\ employs,
  under consideration of the spectral range to be fit, AI to eliminate near
  redundancies in the SSP base and increase computational efficiency.
 
\end{itemize}

The second module in \FADO\ deals with the fitting procedure. This is to be
considered the core component of the code, which uses genetic DEO (see
Sect.~\ref{FADO:DEO} for details), a numerical approach that is best-suited
for multi-objective optimization and is being applied for the first time in
population spectral synthesis:

\begin{itemize}

\item[$\psi$1)] {\rem{On-the-fly measurement of emission line fluxes and
    EWs}}: At first stage in each iterative loop, emission-line fluxes and EWs
  are measured and checked with respect to their quality\footnote{This
    procedure is based on a set of prescriptions originally integrated in the
    PORTO3D pipeline \citep{P13,Gomes2016}, which has been extensively applied
    to the analysis of CALIFA integral field spectroscopy data.}.  The quality
  control first involves a sequential check of various quantities and their
  errors inferred from DEO-based Gaussian line fitting and deblending,
  such as the full width at half maximum (FWHM) and the difference between the
  central wavelength $\delta\lambda_0$ of emission lines. This step is meant
  to identify as outliers spurious line features (e.g., residuals from
  the cosmics or sky correction, noise peaks) on the basis of their FWHM,
  $\delta\lambda_0$, or large uncertainties. For instance, the \ha+[N{\sc ii}]
  deblending solution would be rejected if the [N{\sc ii}]$_{6548/6584}$ lines
  differ in their FWHM by more than an error-dependent tolerance bound, or in
  the case that the redshift-corrected $\delta\lambda_0$'s between the
  \ha\ and [N{\sc ii}]$_{6548/6584}$ lines do not match the nominal value.
  Subsequently, this routine checks whether various emission-line ratios
  (e.g., between hydrogen Balmer lines, or the [S{\sc ii}] 6717/6731 line
  ratio used for the determination of the electron density) fall within the
  range of theoretically expected values.  As an example, the \ha+[N{\sc ii}]
  deblending would be considered unsuccessful if the [N{\sc ii}] 6584/6548
  ratio deviates from the nominal value of $\approx$3, or the [S{\sc ii}]
  6717/6731 flux ratio is lower than $\approx$0.45, which would imply an
  abnormally high electron density ($> 10^4$ cm$^{-3}$) for a SF galaxy.
  \FADO\ in its current (v.1) version approximates emission lines with
  single-Gaussian profiles, which means that does not allow for line
  decomposition into multiple components differing in their central wavelength
  and FWHM \citep[as observed in some starburst galaxies,
    e.g.,][]{Amorin2012b}; this option, besides provision for other fitting
  functions (e.g., Lorentzian, Gauss-Hermite) is foreseen in future releases
  of the code.

\item[$\psi$2)] {\rem{Determination of the physical conditions in the gas}}:
  Determination of the electron density n$_{\rm e}$, temperature T$_{\rm e}$
  and extinction A$_{V,{\rm neb}}$ in the nebular component.  Whenever an
  accurate determination of n$_{\rm e}$ and T$_{\rm e}$ is impossible (as, for
  example, in the case of a virtually lineless passive galaxy,
  cf. Fig.~\ref{fig:fit_Passive}) \FADO\ assumes standard conditions ($10^2$
  cm$^{-3}$ and $10^4$ K, respectively).

\item[$\psi$3)] {\rem{Decision-tree based choice of fitting strategy and
    convergence schemes}}: By default the fitting scheme of \FADO\ aims at
  consistency between observed and predicted line fluxes, EWs (only hydrogen
  Balmer lines in the current version) and nebular continuum (see
  Sect.~\ref{FADO:DEO}\&\ref{FADO:PF} for details). However, there are
  currently three (3) different fitting schemes that can be chosen by the user
  or are automatically set by \FADO\ during execution of block $\alpha$, or
  subsequently in this module, depending on the characteristics of the input
  spectrum and the quality check under $\psi$1.  For instance, in the case of
  unsuccessful or uncertain \ha/[N{\sc ii}] deblending, \FADO\ switches from
  the full-consistency mode (\FCmode) to the partial consistency mode
  (\NCmode), which does not require consistency for Balmer-line fluxes and EWs
  (see Sect.~\ref{FADO:EmLin} for details).

\item[$\psi$4)] {\rem{Multiple evolutionary threads}}: Creation and evolution
  of multiple evolutionary threads (can be seen as multiple Markov Chains in
  the parameter space) converging to solutions (individuals or chromosomes,
  each characterized by a set of parameters (genes)).

\item[$\psi$5)] {\rem{Computation of the Lyman continuum (LyC) output from a
    PV}}: Determination of the total LyC photon rate expected from a stellar
  PV through integration of the SSPs of which it is composed shortwards of
  911.76 ~\AA.

\item[$\psi$6)] {\rem{Computation of predicted Balmer-line luminosities \&
    nebular continuum}}: Computation from the LyC photon rate ($\psi$5) of the
  expected nebular continuum SED and Balmer-line luminosities, assuming case~B
  recombination and taking into account the n$_{\rm e}$ and T$_{\rm e}$
  obtained in $\psi$2.

\item[$\psi$7)] {\rem{Multi-objective optimization \& Pareto solution}}:
  Multi-objective optimization scheme that involves successive generations of
  individuals in combination with a convergence monitoring scheme that ensures
  optimal (Pareto) solutions (see Sect.~\ref{FADO:DEO} for details).

\item[$\psi$8)] {\rem{Estimation of uncertainties in the best-fitting PV}}:
  Once full convergence is reached, and based on the optimum fit and the
  parameter variability within and among individual evolutionary threads
  (cf. $\psi$4), formal uncertainties in all primary quantities that compose
  the best-fitting PV (individual SSP contributions, extinction, velocity
  dispersion) are estimated.

\end{itemize}

The third main module, which post-processes and exports the spectral modeling
output comprises the following components:

\begin{itemize}

\item[$\omega$1)] {\rem{Final measurement of emission-line fluxes \& EWs}}:
  Final determination of emission-line fluxes and EWs, with the latter being
  computed using for the continuum determination the sum of the best-fitting
  stellar + nebular-continuum SED. Uncertainties in line fluxes are propagated
  into uncertainties in A$_{V,{\rm neb}}$ and diagnostic emission-line ratios.

\item[$\omega$2)] {\rem{Computation of secondary evolutionary quantities}}:
  Computation of physical quantities, such as the stellar mass \mstar\ ever
  formed and presently available, the light- and mass-weighted mean stellar
  age and metallicity, the time t$_{1/2}$ when one half of \mstar\ was in
  place, and others.  Formal uncertainties are evaluated for all secondary
  quantities.

\item[$\omega$3)] {\rem{Exportation of the relevant model output}}: This
  module stores the relevant model output in four (4) FITS files and an ascii
  table (see Sect.~\ref{FADO:Output} for details).

\item[$\omega$4)] {\rem{Graphical output}}: Automatically generated graphical
  output (EPS format) with various layout options for the sake of
  visualization and control of the model results.

\end{itemize}

It is worth noting that \FADO\ already integrates a number of additional
features that are expected to be fully implemented and unlocked from its
second version (v.2) onward, such as, provision for the estimation of the LyC
photon escape fraction and for supplying the spectrograph's line spread
function for the sake of improved kinematical fitting. Another option to be
offered in v.2 is the automatic rejection from the SSP library of elements
older than the age of the Universe at the redshift of the galaxy under study,
depending on the assumed cosmology.

\section{The mathematical foundation of \FADO: Multi-objective
  optimization using genetic differential evolution in conjunction with 
  self-consistency requirements\label{FADO:DEO}}
Linear and non-linear optimization problems may be characterized, depending on
their definition and complexity, by multiple local maxima in a
multi-dimensional likelihood topology.  This makes the performance of any
stochastic code unsatisfactory or even intractable under certain conditions
(e.g., increase of number of variables or observables). For bounded-value
problems with linear/non-linear constraints, the search for feasible regions
in the parameter space is, in most cases, a non-deterministic polynomial-time
hard (NP-hard) problem with mutually conflicting objectives.

Optimization problems can be mathematically formulated in terms of one main
objective or fitness function with equality and inequality constrains:
\smallskip
\begin{equation}
  \begin{split}
    \textrm{\it{minimization of }}F(\vec{p}) , \\
    \textrm{\it{subject to }}A_j(\vec{p}) \le 0,\quad j=1,2,...,\iota , \\
    B_k(\vec{p}) = 0,\quad k=1,2,...,\kappa , \\
    C_m \le p_m \le D_m,\quad m=1,2,...,\mu,
  \end{split}
\end{equation}

\noindent where $\vec{p}$ is the parameter vector, $F(\vec{p})$, $A(\vec{p})$
and $B(\vec{p})$ stand for a set of functions of the parameters and $C_m$ and
$D_m$ are real-value numbers.

In the specific case of population synthesis codes, the main objective
function $F(\vec{p})$ to be minimized is, in general, the $\chi^2$ function:

\begin{equation}
\chi^2 = \sum_\lambda^{N_\lambda} \left[ \frac{(O_\lambda - M_\lambda)}{e_\lambda} \right]^2 ,
\end{equation}

\noindent with $O_\lambda$, $M_\lambda$ and $e_\lambda$ representing the
observed, modeled and flux error spectrum, respectively. The number of
wavelengths is given by $N_\lambda$. The equality and inequality constrains
(given by the $A_j$'s and $B_k$'s) are in general not used, with the exception
of the positively defined fractions for the light contribution of the base
elements in the model.  However, in the few cases where they are used (e.g.,
photometric bands, colors, Lick indices, etc.), the constraining handling
technique to be adopted is the "scalarization" procedure that converts a
multi-objective problem into an one single-scalar function $F^\prime(\vec{p})$
using the weighted-sum of the multiple objectives:

\begin{equation}
\textrm{\it{minimization of }} F^\prime(\vec{p}) = \delta F(\vec{p}) + \sum_j^\iota \alpha_j A_j(\vec{p}) + \sum_k^\kappa
\beta_k B_k(\vec{p}) ,
\end{equation}

\noindent where $\delta$, $\alpha_j$'s and $\beta_k$'s are positive numbers.
The major caveat of this approach is the non-direct correspondence between the
objectives and their weights, which leaves the final weight-choice to the
decision maker (and generally requires an extensive set of empirical tests).
Obviously, this "scalarization problem" leads to a conceptual weakness for any
code because there is no warranty that by adding extra constraints or
(typically conflicting) objectives the code will continue converging into
sensible optimal Pareto solutions\footnote{A given solution is Pareto optimal
  (also called efficient, non-dominated or non-inferior) when no further
  improvements can be made to that solution without any trade-off between the
  objectives. Such a solution is named after Vilfredo Federico Damaso Pareto
  (1848-1923) who has invented this concept in the field of microeconomics.}.

In order to reconcile the main objective of conventional \pss\ (minimization
of $\chi^2$ between observed and fitted SED continuum) with the additional
requirement for consistency between the best-fitting PV and expected nebular
(line and continuum) emission characteristics, we have adopted a
multi-objective scheme for the mathematical programming with a genetic
optimization approach over continuous spaces that is called differential
evolution optimization
\citep[DEO,][]{StornPrice1996,StornPrice1997,Price2005}. This is a powerful
global optimization algorithm, easy to implement, well-suited for parallel
computation and, most importantly, reliably converging at an affordable
expense of computational time. The pseudo-code below illustrates how DEO
works:

\begin{table}[h]
    \centering
    \begin{tabular}{|r|l|}
        \hline \cellcolor[gray]{0.8}  1: &\textbf{BEGIN PROGRAM} \\
        \cellcolor[gray]{0.8}  2: & while (convergence criterion not yet met) $\{$ \\
        \cellcolor[gray]{0.8}  3: & \hspace{0.8cm}// \it{\vec{x}: Current population vector of individuals} \\
        \cellcolor[gray]{0.8}  4: & \hspace{0.8cm}// \it{\vec{y}: New     population vector of individuals} \\
        \cellcolor[gray]{0.8}  5: & \hspace{0.8cm}Generate population vector $\vec{x}=(x_1,...x_N)$ \\
        \cellcolor[gray]{0.8}  6: & \hspace{0.8cm}Evaluate the fitness or objective function for \vec{x}\\
        \cellcolor[gray]{0.8}  7: & \\
        \cellcolor[gray]{0.8}  8: & \hspace{0.8cm}for ($i=0; i<N$; $i$++) $\{$\\
        \cellcolor[gray]{0.8}  9: & \hspace{1.25cm}Randomly choose three individuals from \vec{x} \\
        \cellcolor[gray]{0.8} 10: & \hspace{1.25cm}Breed and generate offspring $y_i$ using DE rules\\
        \cellcolor[gray]{0.8} 11: & \hspace{1.25cm}Compute fitness function for $y_i$ \\
        \cellcolor[gray]{0.8} 12: & \hspace{1.25cm}if ( F($y_i$) $\leq$ F($x_i$) ) $\{$ // \it{Choose best suited} \\
        \cellcolor[gray]{0.8} 13: & \hspace{1.55cm}Replace $x_i$ by $y_i$\\
        \cellcolor[gray]{0.8} 14: & \hspace{1.25cm}$\}$ \\
        \cellcolor[gray]{0.8} 15: & \hspace{0.8cm}$\}$\\
        \cellcolor[gray]{0.8} 16: & $\}$\\
        \cellcolor[gray]{0.8} 17: &\textbf{END PROGRAM} \\ \hline
    \end{tabular}
\end{table}
 
Furthermore, we have implemented a constraint violation method
\citep[e.g.,][]{CoelhoCoelho2000,Azad2013} involving feasibility and dominance
rules for individuals (solutions) in order to circumvent the scalarization
problem (see above) and give to the code the flexibility to accept additional
constraints at a minimum expense in CPU time.  These prescriptions are as
follows:

\begin{itemize}
\item[a)] Individuals satisfying the defined constraints are given a higher
  "survival" probability over non-feasible ones
\item[b)] Among two individuals within a feasible region of the parameter
  space, the one having better fitness survives
\item[c)] Among two individuals within an unfeasible region, the one that
  violates a lower number of constrains survives.
\end{itemize}

Therefore, the objective or fitness function of an individual can be written as:

\begin{equation}
\Phi(\vec{p}) =
\begin{cases}
   F(\vec{p}),& \text{if the individual is in a feasible region}\\
   F^{\rm max}(\vec{p})+\zeta(\vec{p}),                                & \text{otherwise}
\end{cases}
\end{equation}

\noindent where $F^{\rm max}(\vec{p})$ and $\zeta(\vec{p})$ are the worst
fitness value of all individuals and the constraint violation function,
respectively.

For a better performance, we also modified the DEO algorithm in order to
incorporate an extra feature for computing probabilities for the survival of
the fittest (or, conversely, the least adapted) in the offspring generation,
with the probability of the least adapted to survive decreasing as the
evolution progresses.  This has a close analogy to the Simulated Annealing
method \citep{Kirkpatrick1983,Cerny1985} where a "temperature" parameter is
used to regulate the acceptance or rejection of least probable regions in the
parameter space for each iteration. The main objective of this particular
routine is to avoid premature convergence into a local minimum and ensure an
adequate level of gene (parameter) variability throughout the fitting
process. This regulatory probability for the least adapted to survive is
integrated in line~12 of the pseudocode.

Last but not least, the computational efficiency of iterative stochastic
methods may be limited by a typically non-optimal termination criterion, which
might be, for example, a rigid minimum for the number of iterations that
ensure convergence.  This problem becomes even more complex in multi-objective
programming, where potentially conflicting objectives do not permit full
convergence in the parameter space. For this reason, \FADO\ integrates a
refined searching strategy for the Pareto optimal solution, which is based on
a comparison between the variance of a lineage of chromosomes with the
variance of the individuals within a given population (evolutionary thread).
When these quantities are virtually of the same order, or both show little
evolution over several successive generations (iterations), the process is
considered to have converged to the optimum.  A similar statistical method is
used in MCMC procedures making use of the \citet{GR1992} convergence test.

\section{The physical foundation of \FADO\label{FADO:PF}}

The standard model in state-of-the-art \pss that is usually used to fit the
observed SED continuum of galaxies by means of a linear combination of
discrete spectral components (individual stellar spectra or SSPs) can be
expressed as:

\begin{equation}
  F_\lambda = \sum_{j=1} ^{N_\star} M_j L_{j,\lambda} \times 10^{-0.4 A_\lambda} \otimes S(\vec{v_\star},\vec{\sigma_\star} ) ,
  \label{eq:FluxStellarMass}
\end{equation}

\noindent where $M_j$ is the stellar mass, $L_{j,\lambda}$ is the spectrum of
the $j^{\rm th}$ element in units of $[$L$_\odot$ M$_\odot^{-1}$
  \AA$^{-1}$$]$, $A_\lambda$ is the extinction as a function of wavelength and
$S(\vec{v_\star},\vec{\sigma_\star})$ is a kinematical kernel used to convolve
the modeled spectrum in order to simulate the stellar systemic velocity
$\vec{v_\star}$ and velocity dispersion $\vec{\sigma_\star}$ of galaxies.
This equation can be rewritten such as to explicitly quantify the light
fractions $x_{j,\lambda_0}$ of individual spectral components as:

\begin{equation}
\begin{split}
  F_\lambda = F_{\lambda_0} \sum_{j=1} ^{N_\star} x_{j,\lambda_0} \frac{L_{j,\lambda}}{L_{j,\lambda_0}} \times 10^{-0.4 A_V (q_\lambda - q_{\lambda_0})} \otimes S(\vec{v_\star},\vec{\sigma_\star} ) ,\\
   \textrm{\it{subject to }} \sum_j^{N_\star} x_{j,\lambda_0} = 1 , \\
   \textrm{\it{and} } 0 \le x_{j,\lambda_0} \le 1, \forall j ,
  \end{split}
\label{eq:FluxStellarLight}
\end{equation}

\noindent where $F_{\lambda_0}$ is the flux at the normalization wavelength
$\lambda_0$, $q_\lambda$ is the ratio of $A_\lambda$ over the $V$-band
extinction $A_V$ that is inferred from an assumed reddening law, and
$L_{j,\lambda_0}$ is the luminosity at $\lambda_0$ of the $j^{\rm th}$ base
element.  The spectral fitting is generally done using light contributions,
which are then converted into mass fractions.  Comparison of
Eqs.~\ref{eq:FluxStellarMass} and \ref{eq:FluxStellarLight} shows that the
conversion factor between light and mass reads as:

\begin{equation}
 M_j = \frac{x_{j,\lambda_0}}{L_{j,\lambda_0}} \times F_{\lambda_0} \times 10^{+0.4 A_V q_{\lambda_0}} .
\end{equation}

\smallskip
As already pointed out in Sect.~\ref{FADO_in_a_nutshell}, a key innovation of
\FADO\ over currently available \pss\ codes is the implementation of the
contribution by nebular continuum and hydrogen recombination lines. Therefore,
Eq.~\ref{eq:FluxStellarMass} for the total continuum now includes an extra
term that takes into account the nebular continuum:

\begin{equation}
 \begin{split}
  F_\lambda =  \overbrace{ \sum_{j=1} ^{N_\star} M_{j,\lambda_0} L_{j,\lambda} \times 10^{-0.4 A_V q_\lambda} \otimes S(\vec{v_\star},\vec{\sigma_\star} )}^{STELLAR} \\
  +\\
  \underbrace{ \Gamma_\lambda({\rm n}_{\rm e},{\rm T}_{\rm e}) \times 10^{-0.4 A_V^{\rm neb} q_\lambda} \otimes N(\vec{v_\eta},\vec{\sigma_\eta} )}_{NEBULAR} ,
  \end{split}
  \label{eq:FluxSteNebMass}
\end{equation}

\noindent where $A_V^{\rm neb}$ is the gas extinction,
$N(\vec{v_\eta},\vec{\sigma_\eta})$ is the convolution kernel used to take
into account the nebular kinematics and $\Gamma_\lambda({\rm n}_{\rm e},{\rm
  T}_{\rm e})$ is the nebular continuum spectrum computed assuming that all
LyC photons ($\lambda \leq 911.76 ~\AA$) from stellar populations are absorbed
and re-emitted at longer wavelengths. For case~B recombination,

\begin{equation}
\Gamma_\lambda({\rm n}_{\rm e},{\rm T}_{\rm e}) = \gamma^{\rm eff} \frac{c}{\lambda^2 \alpha_B({\rm H},{\rm n}_{\rm e},{\rm T}_{\rm e})}\textrm{LyC in units of }[\textrm{L}_\odot \textrm{ } \AA^{-1} ]
\end{equation}

\noindent the terms in the equation above correspond to the total
recombination coefficient $\alpha_B({\rm H},{\rm n}_{\rm e},{\rm T}_{\rm e})$
from hydrogen \citep{Osterbrock2006} and the effective continuous emission
coefficient $\gamma_{\rm eff}$ computed following the standard procedure of
combining the contributions from H and He (other species do not significantly
contribute to the continuum emission and can be neglected, see
\citet{Brown1970} for a full explanation), which depends on the 2-photon
\citep{Nussbaumer1984}, free-free and free-bound \citep{Brown1970,ES06}
emission. The $\gamma_{\rm eff}$ coefficient can be expressed as:

\begin{equation}
\gamma_{\rm eff} = \gamma(\textrm{HI}) + \gamma(2q) +
\frac{n(\textrm{He}^+)}{n(\textrm{H}^+)}\gamma(\textrm{HeI})+\frac{n(\textrm{He}^{++})}{n(\textrm{H}^+)}\gamma(\textrm{HeII})$$ ,
\end{equation}

\noindent where $n(\textrm{H}^+)$ is the density of once-ionized hydrogen and
$n(\textrm{He}^+)$ and $n(\textrm{He}^{++})$ that of the once and twice
ionized helium.  $\gamma(\textrm{H{\sc i}})$, $\gamma(\textrm{He{\sc I}})$ and
$\gamma(\textrm{He{\sc ii}})$ are the total emission coefficients of H{\sc i},
He{\sc i} and He{\sc ii} taking into account free-free and free-bound emission
processes.

As mentioned in Sect.~\ref{FADO1-main}, n$_{\rm e}$ and T$_{\rm e}$ are
computed on the fly in module $\psi$2.  To this end, \FADO\ uses a five-level
photoionization model based on the same principles as the iterative code FIVEL
\citep{deRobertis1987-FIVEL,ShawDufour1994,ShawDufour1995}.  Starting from an
initial estimate of n$_{\rm e}$ and T$_{\rm e}$ based, respectively, on the
[S{\sc ii}] 6717/6731 and [O{\sc iii}] 4959+5007/4363 line ratios, this
routine solves iteratively for (n$_{\rm e}$,T$_{\rm e}$), reaching convergence
in typically 3 iterations. Whenever the determination of (n$_{\rm e}$,T$_{\rm
  e}$) from these line ratios or alternative diagnostics (e.g., [N{\sc i}]
5200/5198 and [O{\sc ii}] (3726+3729)/(7320+7330), respectively) fails,
standard conditions (n$_{\rm e}$=10$^2$ cm$^{-3}$,T$_{\rm e}$=10$^4$ K) are
assumed.  \FADO\ (v.1) also offers the option of n$_{\rm e}$ and T$_{\rm e}$
being fixed by the user.
 
The nebular extinction $A_V^{\rm neb}$ is computed from the Balmer decrement
whenever it can be reliably determined from the net emission-line spectrum and
is used to constrain the spectral fit.

Since \pss\ techniques attempt to match the observed spectral continuum with a
synthetic SED, it is convenient to express Eq.~\ref{eq:FluxSteNebMass},
subject to similar constraints as Eq.~\ref{eq:FluxStellarLight}, in terms of
light fractions:

\begin{equation}
 \begin{split}
  F_\lambda =  F_{\lambda_0} \sum_{j=1} ^{N_\star} x_{j,\lambda_0} \frac{L_{j,\lambda}}{L_{j,\lambda_0}} \times 10^{-0.4 A_V (q_\lambda - q_{\lambda_0})} \otimes S(\vec{v_\star},\vec{\sigma_\star} ) \\
  +\\
  F_{\lambda_0} y_{\lambda_0} \frac{\Gamma_\lambda({\rm n}_{\rm e},{\rm T}_{\rm e})}{\Gamma_{\lambda_0}({\rm n}_{\rm e},{\rm T}_{\rm e})} \times 10^{-0.4 A_V^{\rm neb} (q_\lambda - q_{\lambda_0})} \otimes N(\vec{v_\eta},\vec{\sigma_\eta} ) ,\\
  \textrm{\it{subject to }} \sum_j^{N_\star} x_{j,\lambda_0} + y_{\lambda_0} = 1 , \\
  \textrm{\it{and} } 0 \le (x_{j,\lambda_0},y_{\lambda_0}) \le 1, \forall j ,
  \end{split}
  \label{eq:FluxSteNebLight}
\end{equation}

The fractional contribution $y_{\lambda_0}$ of the nebular continuum at
$\lambda_0$ is in the full self-consistency fitting mode ({\rem FCmode};
Sect.~\ref{FADO:EmLin}) constrained by the PV with $\vec{x_{\lambda_0}} =
(x_{1,\lambda_0},\ldots,x_{N_\star,\lambda_0})$, which contains the fractional
light contributions of SSPs of a given age and metallicity. The total number
of LyC photons is the linear sum of all ionizing components weighted by mass
($\sum_j M_j $ LyC$_{j}$) and, since $y_{\lambda_0}$ is coupled with the SFH:

\begin{equation}
\begin{split}
y_{\lambda_0} & = & \sum_j^{N_\star} \frac{x_{j,\lambda_0}}{L_{j,\lambda_0}} \frac{\textrm{ LyC}_{j}}{\textrm{LyC}} \textrm{ } \Gamma_{\lambda_0} ({\rm n}_{\rm e},{\rm T}_{\rm e}) \times 10^{-0.4 q_{\lambda_0} \overbrace{(A_V^{\rm neb}-A_V)}^{\textrm{Nebular - Stellar}} }\\
& = & \frac{1}{F_{\lambda_0}} \sum_j^{N_\star} M_j \frac{\textrm{ LyC}_{j} \textrm{ }}{\textrm{LyC}} \Gamma_{\lambda_0} ({\rm n}_{\rm e},{\rm T}_{\rm e}) \times 10^{-0.4 q_{\lambda_0} A_V^{\rm neb}} \\ & = & \underbrace{\frac{1}{F_{\lambda_0}}\textrm{ } \Gamma_{\lambda_0} ({\rm n}_{\rm e},{\rm T}_{\rm e})}_{\textrm{Nebular to Total at }\lambda_0} \times 10^{-0.4 q_{\lambda_0} A_V^{\rm neb}} ,
\end{split}
\end{equation}
with the (n$_{\rm e}$,T$_{\rm e}$) obtained in module $\psi$2 (Sect.~\ref{FADO1-main}).

We note that the treatment of nebular emission in \FADO\ (v.1) on the basis of
the standard assumption of case~B recombination has been kept at a level of
simplicity that allows for computation of only those nebular characteristics
that are self-consistently reproduced by the model. Specifically, since the
solution is driven by the requirement for consistency between the best-fitting
PV and the observed nebular continuum SED and, additionally, hydrogen
Balmer-line luminosities (\NCmode\ and \FCmode, respectively;
cf. Sect.~\ref{FADO:EmLin}) it is both unnecessary and computationally
expensive to internally track the full nebular line spectrum of a galaxy (for
instance, forbidden lines, which would require assumptions on, for
  example, the ionization parameter and geometry)\footnote{Even though this
  is beyond the main scope of \FADO, future releases of the code will
  optionally facilitate post-processing of the best-fitting PV with
  dedicated photoionization codes (e.g., Cloudy, \textcolor{blue}{Ferland et
    al. 2013}; Cloudy3D, \textcolor{blue}{Morisset \& Stasi\'nska 2006,2008};
  Mappings~1e, \textcolor{blue}{Binette et al. 2012}) for the sake of detailed
  modeling of the full emission-line spectrum.}. It should be born in mind
that, given the much greater mathematical and computational challenges of
\pss\ (the inverse approach), as compared to the forward (\ess)
approach, already the implementation and self-consistent treatment of nebular
(continuum and Balmer-line) emission is a major improvement over
state-of-the-art \pss\ codes.  On the other hand, the versatile DEO-based
architecture of \FADO\ (cf. Sect.~\ref{FADO:DEO}) might, in principle, be able
to accommodate further constraints coupled with a more detailed treatment of
the nebular component (e.g., forbidden lines, interstellar metallicity, shock
excitation and radiation transfer, multiple electron temperatures,
non-equilibrium ionization) at a reasonable expense in computational
time. This possibility will be explored in the course of the further
development of the code.

\section{Spectral fitting modes in \FADO \label{FADO:EmLin}}

\FADO\ integrates different fitting modes with the provision of auto-switching
between them, depending on the spectral classification and emission-line
quality control in modules $\alpha$5 and $\psi$3:

\begin{itemize}
\item[1-]\FCmode\ ({\bf F}ull-{\bf C}onsistency {\bf mode}): Spectral modeling
  aiming at consistency between observed and predicted SED continuum (stellar
  plus nebular) and Balmer emission-line luminosities and EWs.

\item[2-]\NCmode\ ({\bf N}ebular-{\bf C}ontinuum {\bf mode}): as \FCmode\ but
  without the requirement for consistency between predicted and observed
  Balmer-line luminosities and EWs. Notwithstanding this fact, the requirement
  of fitting the observed continuum, in particular around the Balmer and
  Paschen jump, entails in itself a "soft" consistency condition.

\item[3-]\STmode\ ({\bf ST}ellar {\bf mode}): State-of-the-art \pss\ modeling
  with purely stellar SSP templates, that is, neglecting nebular
  continuum emission and without the requirement of consistency of the
  best-fitting PV with the observed nebular emission characteristics.
\end{itemize}

In its first public release, the code implements a conservative approach that
allows for fitting in the \FCmode\ only for galaxies being classified as
SF/composite on the basis of classical BPT diagnostics. Due to the still
controversial origin of LINER ionization \citep[e.g.,][and references
  therein]{Binette1994,Allen08,Ho08,SharpBlandHawthorn10,P13,Gomes2016,Herpich16},
\FADO\ (v.1) adopts for such systems by default the \NCmode. As for Seyfert
galaxies, the \FCmode\ is automatically dropped if spectral fitting is
attempted with purely stellar SSPs (i.e., the base library lacks a
power-law component as a proxy to the featureless continuum of an AGN). We
note that \FADO\ allows the user to disable these cautionary measures in its
configuration file, even though this practice is not recommended.

The on-the-fly spectroscopic classification scheme in modules $\alpha$5 \&
$\psi$3 currently uses four (4) emission-lines involved in classical BPT
diagrams in conjunction with a 2D probability density function treatment
(multivariate normal distribution): a point lying on the BPT diagram with an
associated error on the log [O{\sc iii}]/\hb\ and log [N{\sc ii}]/\ha\ plane
is assigned with probabilities of being classifiable as AGN (Seyfert), LINER,
composite and SF galaxy photoionized by OB stars. The respective loci of these
four spectroscopic classes are based on demarcation lines from
\citet[][]{Stasinska2008}, \citet[][]{Kauffmann2003c}, \citet[][]{Kewley2001}
and \citet[][]{Schawinski2007}.  The main equation used assumes that errors in
log[O{\sc iii}]/\hb\ and log[N{\sc ii}]/\ha\ are purely Gaussian:

\begin{equation}
\begin{split}
f(\vec{x}) = \frac{1}{\sqrt{(2\pi)^k|\Sigma|}} e^{ -\frac{1}{2}(\vec{x}-\vec{\xi})^T \Sigma^{-1}(\vec{x}-\vec{\xi})},
\end{split}
\end{equation}

\noindent and uncorrelated, where the vector $\vec{x}$ $\equiv$ $(
\textrm{x$_1$:}\log [\textrm{O{\sc iii}}]/\textrm{\hb}$ , $\textrm{x$_2$:}\log
       [\textrm{N{\sc ii}}]/\textrm{\ha})$, $\Sigma$ is the symmetric
       covariance matrix, $k$ is the dimension of the multivariate Gaussian
       (in our case equal 2) and $\xi$ is the mean value vector. Therefore,
       the elliptical Gaussian function corresponding to $\sigma_{x_1}$ and
       $\sigma_{x_2}$ is given by:

\begin{equation}
f(x_1,x_2) = \frac{1}{2 \pi \sigma_{x_1} \sigma_{x_2}} e^{-\left( \frac{[(x_1-\xi_{x_1})^2}{2\sigma_{x_1}^2}+\frac{(x_2-\xi_{x_2})^2}{2\sigma_{x_2}^2} \right)}.
\end{equation}

As an example, we show in Fig.~\ref{fig:BPT_WHAN} three emission-line spectra
processed by \FADO\ (see Sect.~\ref{FADO:three} for a full discussion) lying
on the SF (\object{CGCG 007-025}), composite (2MASX J154552094+0812244) and
LINER/retired (\object{MCG +00-03-030}) regimes with $\log [\textrm{N{\sc
      ii}}]/\textrm{\ha}$ = $(-1.933,-0.201,0.117)$, $\log [\textrm{O{\sc
      iii}}]/\textrm{\hb}$ = $(0.754,-0.060,0.229)$, and the corresponding 2D
probabilities $\pi$ in percentage for the respective galaxies falling in the
class of SF (100.0,1.26,0.76), composite (0.0,94.40,16.90), LINER
(0.0,3.71,53.53) and Seyfert (0.0,0.63,28.81).  Future releases of \FADO\ will
include covariance terms for other emission-line diagnostics \citep[e.g.,
  after][or alternatives, such as the WHAN classification by
  \citealt{CidFernandes2010,CidFernandes2011}]{Veilleux1987} for the sake of a
more refined spectral classification with 2D probability functions.


\section{Primary and secondary output quantities from spectral modeling with \FADO\ \label{FADO:Output}}

This section provides a concise description of some of the primary and
secondary output quantities from \FADO\ and their storage in four (4)
different output Flexible Image Transport System (FITS) files, each of them
containing in its header a detailed description of its content.

\begin{itemize}
\item[{\rem i)}] {\rem 1D spectra (\texttt{\textunderscore}1D):} This file
  holds the wavelength-dependent quantities, specifically the
  de-redshifted and rebinned observed spectrum, the error spectrum, a mask
  with spurious features and the best-fitting synthetic SED along with the
  average, median and standard deviation of the solutions obtained from the
  multiple individuals involved in the fit. The stellar and nebular SED (in
  the case of Seyfert spectra, additionally the power-law component of the
  AGN) composing the best-fitting spectrum, the line-spread function
  (independent on $\lambda$ for \FADO\ v.1) and other 1D arrays are also
  stored in this file.\\

\item[{\rem ii)}] {\rem Statistics (\texttt{\textunderscore}ST):} To
  facilitate statistical analysis of large galaxy samples, it is useful to
  reduce the \pss\ output into a few key variables characterizing the stellar
  and nebular component of a galaxy. These secondary model quantities,
  computed in module $\omega$2, include the first moments of the best-fitting
  PV, that is, the mean stellar age and metallicity, expressed both in
  linear and logarithmic form:

\begin{equation}
\begin{split}
\langle t_\star \rangle_{\cal L} &= \sum_j^{N_\star} \frac{L_{j,\lambda_0}}{L^{\rm total}_{\lambda_0}} t_j &=& \sum_j^{N_\star} x_{j,\lambda_0} t_j &\textrm{ weighted by light,}\\
\langle t_\star \rangle_{\cal M} &= \sum_j^{N_\star} \frac{M_j}{M_\star} t_j                         &=& \sum_j^{N_\star} \mu_j t_j &\textrm{ weighted by mass,}
\end{split}
\end{equation}
\noindent and the second as:
\begin{equation}
\begin{split}
\langle \log t_\star \rangle_{\cal L} &= \sum_j^{N_\star} \frac{L_{j,\lambda_0}}{L^{\rm total}_{\lambda_0}} \log t_j &=& \sum_j^{N_\star} x_{j,\lambda_0} \log t_j ,\\
\langle \log t_\star \rangle_{\cal M} &= \sum_j^{N_\star} \frac{M_j}{M_\star} \log t_j                         &=& \sum_j^{N_\star} \mu_j \log t_j ,
\end{split}
\end{equation}

\noindent with $L^{\rm total}_{\lambda_0}$ standing for the total luminosity
of the galaxy at the normalization wavelength $\lambda_0$, $t_j$ and $M_j$
being the age and mass presently available in the $j^{\textrm{th}}$ SSP
contributing a mass fraction $\mu_j$ to \mstar.  Other terms have their usual
meaning as previously defined.

The same formalism can be used for the mean stellar metallicity:
\begin{equation}
\begin{split}
\langle Z_\star \rangle_{\cal L} &= \sum_j^{N_\star} \frac{L_{j,\lambda_0}}{L^{\rm total}_{\lambda_0}} Z_j &=& \sum_j^{N_\star} x_{j,\lambda_0} Z_j &\textrm{ weighted by light,}\\
\langle Z_\star \rangle_{\cal M} &= \sum_j^{N_\star} \frac{M_j}{M_\star} Z_j                         &=& \sum_j^{N_\star} \mu_j Z_j &\textrm{ weighted by mass,}
\end{split}
\end{equation}
\noindent and its logarithmic form:
\begin{equation}
\begin{split}
\langle \log Z_\star \rangle_{\cal L} &= \sum_j^{N_\star} \frac{L_{j,\lambda_0}}{L^{\rm total}_{\lambda_0}} \log Z_j &=& \sum_j^{N_\star} x_{j,\lambda_0} \log Z_j ,\\
\langle \log Z_\star \rangle_{\cal M} &= \sum_j^{N_\star} \frac{M_j}{M_\star} \log Z_j                         &=& \sum_j^{N_\star} \mu_j \log Z_j .
\end{split}
\end{equation}
Besides the first moments of these distributions, several other quantities
also exported, including the ever formed and presently available stellar mass
({\rem M}$^{\rm e}$ and {\rem M}$^{\rm p}$, respectively) the mass fraction of
stars younger (and older) than 100 Myr, 1 Gyr and 5 Gyr, the rate of LyC
photons being capable of ionizing hydrogen ({\rem Q}$_{\rm H}$) and
singly/double ionizing helium ({\rem Q}$_{\rm HeI}$,{\rem Q}$_{\rm HeII}$),
among others. \\

\item[{\rem iii)}] {\rem Emission-Lines (\texttt{\textunderscore}EL):} A
  listing of measured fluxes and EWs, including uncertainties, for currently
  51 lines in the spectral range between 3425 \AA\ and 8617 \AA, ([O{\sc
      ii}]$_{3727,3729}$, H$\delta$, H$\gamma$ [O{\sc iii}]$_{4363}$, \hb,
  [O{\sc iii}]$_{4959}$, [O{\sc iii}]$_{5007}$, [N{\sc ii}]$_{5755}$, [O{\sc
      i}]$_{6300}$, [N{\sc ii}]$_{6548}$, \ha, [N{\sc ii}]$_{6584}$, [S{\sc
      ii}]$_{6716}$, [S{\sc ii}]$_{6731}$ among others).\\

\item[{\rem iv)}] {\rem Population Vector (\texttt{\textunderscore}DE):} The
  full population vector (i.e., the light- and mass contributions
  ($x_j,\mu_j$) of SSPs, kinematical parameters, stellar and nebular
  extinction) for all individuals.
\end{itemize}

Additionally, in the case of batch execution of \FADO\ on large data sets, a
listing of the most relevant model quantities is exported in an ascii table.

\section{Some illustrative applications of \FADO\ on galaxy spectra
\label{FADO:three}}
\begin{figure*}
\includegraphics[trim={0cm 15cm 0cm 3cm},clip,scale=1.0,angle=0, width=18cm]{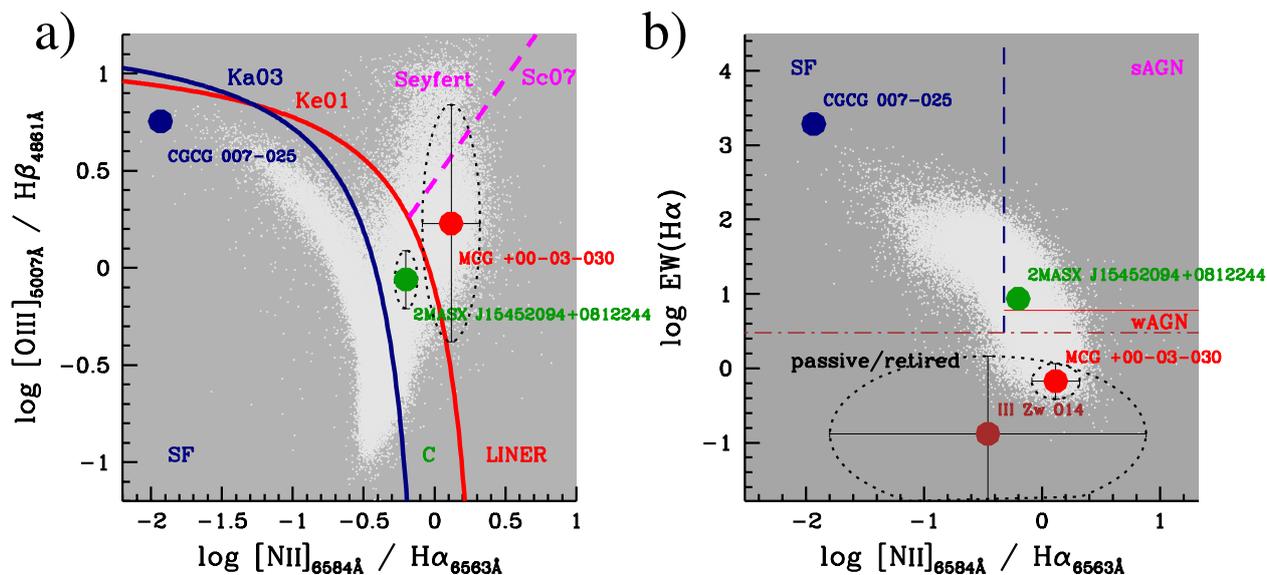}
\put(-16.8,7.5){\color{black}\huge a)}
\put(-8.8,7.5){\color{black}\huge b)}
\caption[]{{\rem a)} Loci of the three emission-line spectra (star-forming SF:
  \object{CGC 007-025}, composite (C): \object{2MASX J15452094+0812244} and
  LINER: \object{MCG +00-03-030}) fitted by
  \FADO\ (Figs.~\ref{fig:fit_SF}-\ref{fig:fit_ETG}) on the BPT diagnostic
  diagram. The error bars and the ellipses depict 1$\sigma$ level
  uncertainties in the emission-line ratios. The demarcation lines correspond
  to the SF/H{\sc ii} regime \citep[blue;][]{Kauffmann2003c} and AGN
  \citep[red;][]{Kewley2001}. The division between Seyfert and LINERs is
  delineated by the magenta dashed line \citep[][]{Schawinski2007}. {\rem{b)}}
  WHAN diagram \citep{CidFernandes2010,CidFernandes2011} with the position of
  the four spectra fitted by \FADO\ in Sect.~\ref{FADO:three} are shown for
  the sake of comparison with the BPT classification. This alternative
  diagnostic diagram attempts to classifiy galaxies in SF, strong AGN (sAGN),
  weak AGN (wAGN) and passive/retired. In this diagram, we additionally
  include the tentative locus of the passive (lineless) galaxy \object{III Zw
    014}, for which BPT emission-line ratios are unavailable.  The white
  shaded area in both diagrams depicts the distribution of galaxies from
  SDSS.}
\label{fig:BPT_WHAN}
\end{figure*}

This section provides illustrative examples of spectral modeling with
\FADO\ of galaxy spectra that are classifiable by BPT or WHAN diagnostics as
star-forming, composite, LINER/retired and passive
(cf. Fig. \ref{fig:BPT_WHAN}). These examples are meant to outline the
functionality and typical graphical output from the code.

\begin{figure*}[t]
\begin{picture}(18.4,13.8)
\put(0.0,0.0){\includegraphics[scale=1.0, angle=0, width=19cm]{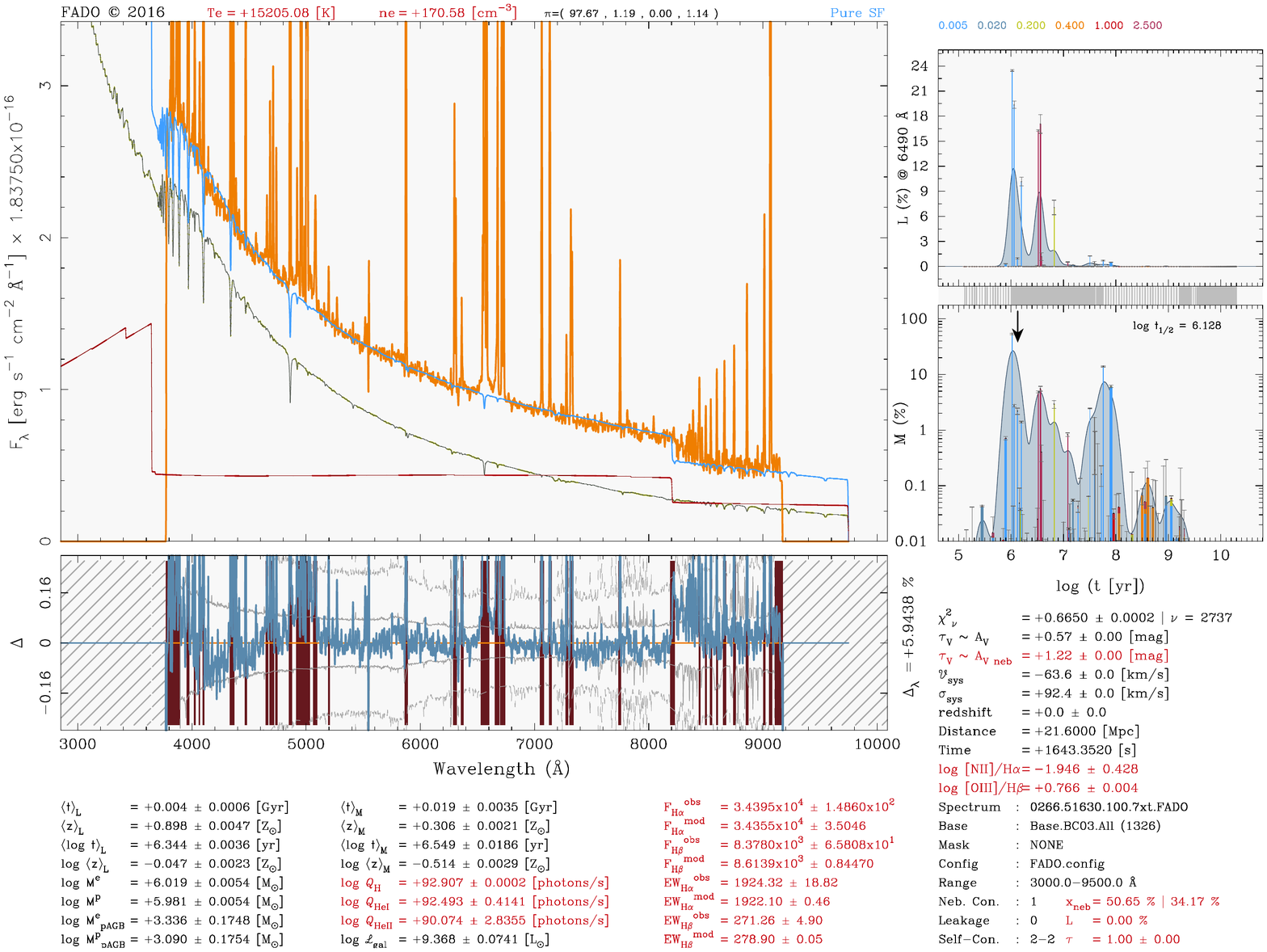}}
\PutLabel{1.6}{13.1}{\color{black}\huge a)}
\PutLabel{14.1}{12.7}{\color{black}\huge c)}
\PutLabel{1.6}{5.4}{\color{black}\huge b)}
\PutLabel{14.1}{9.0}{\color{black}\huge d)}
\end{picture}
\caption[]{{\bf a)} Example of a spectral fit with \FADO\ of the star-forming
  region {\rem a} in the low-metallicity BCD \object{CGCG 007-025}
  \citep[cf. e.g.,][]{Guseva2007}. The SDSS spectrum of the source, corrected
  for Galactic extinction (orange curve), reveals intense nebular emission
  with a \ha\ equivalent width of $\sim$2000 \AA\ and strong contamination by
  nebular continuum emission, as apparent from the complete absence of stellar
  absorption features and the visible Paschen jump at 8207 \AA. The
  best-fitting synthetic SED (open-blue) is composed of stellar and nebular
  continuum emission (dark gray and red, respectively). {\bf b)} Residuals
  between fit and observed spectrum, with the shaded area and the dashed curve
  delineating, respectively, the $\pm$1$\sigma$ and $\pm$3$\sigma$ error
  spectrum. {\bf (c)} Luminosity fraction at the normalization wavelength
  (6490 \AA) and {\bf (d)} stellar mass fraction of the SSPs composing the
  best-fitting population vector as a function of their age. The color-coding
  depicts the metallicity and the vertical bars $\pm$1$\sigma$
  uncertainties. The thin-gray vertical lines connecting both diagrams
  correspond to the ages of the SSPs in the used library. The light-blue
  shaded area in both panels shows an Akima-smoothed \citep{Akima1970} version
  of the SSP contributions, giving in panel d a schematic illustration of the
  star formation history. The electron temperature T$_{\rm e}$ and density
  n$_{\rm e}$ computed by \FADO\ from emission lines label the upper part of
  panel a, together with the probability $\pi$ of the spectrum to fall in the
  locus of SF, composite, LINER and Seyfert galaxies in BPT diagrams. Details
  on the meaning of the labels beneath the four panels are provided in the
  text. Panels c\&d are shown without errors bars for the sake of better
  visibility.}
\label{fig:fit_SF}
\end{figure*}

Figure~\ref{fig:fit_SF} shows a spectral fit to the SF region {\rem a} of the
nearby (D=21.6 Mpc) BCD \object{CGCG 007-025} \citep[cf. e.g.,][hereafter
  G07]{Guseva2007}. The presence of strong ongoing SF activity in this system
is apparent from its SDSS spectrum (orange color in panel {\rem a}), which
shows a plethora of intense narrow nebular emission lines superimposed on a
blue continuum with the Paschen discontinuity (8207 \AA) visible in its red
part. The best-fitting SED (open-blue) comprises stellar and nebular continuum
emission (dark gray and red, respectively), with the latter contributing
$\sim$50\% of the monochromatic luminosity at $\sim$7400 \AA\ and nearly 35\%
of the total luminosity of the modeled SED between 2750 \AA\ and 9750 \AA.
Panel {\rem b} shows the residuals between the observed and modeled spectrum,
with the shaded area and the dashed lines delineating, respectively, the
$\pm$1$\sigma$ and $\pm$3$\sigma$ error spectrum, as automatically determined
by \FADO\ (cf. Sect.~\ref{FADO1}, $\alpha$4).

The spectral fit has been computed in the {\sl full-consistency mode}
(\FCmode; {\rem Self-Con. 2}) and assuming no LyC photon escape ({\rem
  Leakage: 0}) with a library of 1326 SSPs from \citet{BruzualCharlot2003}
spanning a range in age between 1~Myr and 13.75 Gyr for six stellar
metallicities (1/200, 1/50, 1/5, 1/2.5, 1 and 2.5 $Z_{\odot}$) for Padova 1994
evolutionary tracks
\citep{Alongi1993,Bressan1993,Fagotto1994a,Fagotto1994b,Fagotto1994c,
  Girardi1996}, STELIB stellar library \citep{LeBorgne2003} and the
\citet{Chabrier2003} IMF with a lower and upper mass cutoff of 0.1 and 100
M$_{\odot}$. That \FADO\ has reached convergence in the \FCmode\ (2) can be
read off the second number that follows the label {\rem Self-Con}
(lower-right), with the first one corresponding to the fitting mode initially
chosen by the user (in this case 2, which is the default setting for
SF/composite galaxies).  The production rate of LyC photons capable of
ionizing hydrogen ({\rem Q$_{\rm H}$}) and singly/doubly ionizing helium
({\rem Q$_{\rm HeI}$}, {\rem Q$_{\rm HeII}$}) that is predicted by the
best-fitting PV is listed in the second column from bottom-left, and the third
column gives a comparison between the observed and predicted (superscript
{\rem obs} and {\rem mod}, respectively) \ha\ and \hb\ fluxes (in $10^{-17}$
\uflux) and equivalent widths (in \AA).  It is worth noting that the
predicted Balmer-line luminosities and nebular continuum are computed by
taking into account the electron temperature and density ({\rem T$_{\rm
    e}$}=15205 K and {\rem n$_{\rm e}$}=171 cm$^{-3}$, respectively;
cf. labels at the upper part of panel {\rem a}) that have been derived from
the observed spectrum in sub-module $\psi$2a.  It can be seen that the model
yields a good match to the observations, with regard to both the SED
continuum (including the Paschen jump) and the \ha\ and \hb\ flux and EW. We
note that the [O{\sc iii}]-based electron temperature and EW(\hb) inferred
above are in reasonably good agreement with those from longslit spectroscopy
by G07 (15.8 kK and 274 \AA, respectively), which additionally reveals a
strong Balmer jump shortwards of 3646 \AA, in agreement with the SED predicted
by \FADO.

The rightmost panels {\rem c}\&{\rem d} display as a function of age the
luminosity and mass contribution of individual SSPs in the best-fitting PV,
with the color coding depicting their metallicity and the vertical bars
1$\sigma$ uncertainties. Gray vertical lines connecting the diagrams
mark the ages of the SSPs in the used library (eventually, after AI-supported
optimization through sub-module $\alpha$7 in case of a SSP base with more than
800 elements).  The determined systemic velocity $v_{\rm sys}$ (in this and
the following examples, $v_{\rm sys}$ is expressed relative to the recession
velocity given by SDSS) and the intrinsic velocity dispersion ($\sigma_{\rm
  sys}$) are listed in the bottom-right column, along with the $V$-band
extinction obtained for the stellar and nebular component ({\rem A$_{\rm V}$}
and {\rem A$_{\rm V\,neb}$}, respectively) yielding a ratio of $\sim$1/2.
This column additionally lists the determined log([N{\sc ii}]/\ha) and
log([O{\sc iii}]/\hb) ratios, which place the galaxy on the BPT diagram
(Fig.~\ref{fig:BPT_WHAN}) in the locus of star-forming galaxies. The
likelihood ({\rem $\pi$}) in \% for the spectrum being classifiable as SF,
composite, LINER or Seyfert is given by the four numbers on the top-right of
panel {\rem a} (see Sect.~\ref{FADO:EmLin} for details).  Various other
secondary quantities of interest are listed beneath the diagrams, such as the
luminosity- and mass-weighted stellar age and metallicity, in both
linear and logarithmic form (cf. Sect. \ref{FADO:Output}), and the ever formed
and presently available \mstar, both the total one ({\rem M$^{\rm e}$}, {\rem
  M$^{\rm p}$}) and that of the post-AGB (age $\geq$ 100 Myr) stellar
component (subscript {\rem pAGB}). The vertical arrow in the log~t vs M
diagram (panel {\rem d}) marks the age $t_{1/2}$ when 50\% of the present-day
\mstar\ has been in place.

The spectral fit in Fig.~\ref{fig:fit_SF} illustrates several unique
advantages of \FADO\ with special importance to studies of SF galaxies.  One
of them is its ability to converge even in the case of extreme nebular
contamination, while reproducing key spectral features, such as the Paschen
discontinuity and the hydrogen Balmer-line luminosities and EWs.  This is not
the case for spectral fits of SF galaxies with state-of-the-art (purely
stellar) \pss\ codes (e.g., STARLIGHT), for which an in-depth study by CGP17
reveals a strong (typically, one order of magnitude) discrepancy between the
observed Balmer-line luminosities and those implied by the best-fitting
stellar model. This discrepancy obviously also reflects the failure of these
standard \pss\ codes to place meaningful constraints on the recent SFH of
galaxies, consequently also to other evolutionary indicators, such as the sSFR
or recent-to-past SFR.

Spectral modeling with \FADO\ of low-metallicity, high-excitation BCDs, such
as \object{CGCG 007-025} can obviously benefit from inclusion of the Balmer
jump, which future versions of the code will additionally take advantage of in
order to supplement the model output with T$_{\rm e}$ determinations in the
H$^+$ zone \citep[cf. e.g.,][]{Guseva2007} through an upgraded version of
sub-module $\psi$2.  In this context, it is also worth reiterating that
already the nebular continuum in itself (i.e., the pure requirement posed in
NCmode for the synthetic SED to reproduce the Balmer and Paschen jump, even
without explicitly demanding the predicted hydrogen Balmer-line luminosities
and EWs to match the observed ones) implicitly holds constraints that guide
\FADO\ toward consistency between observed and predicted
\ige\ characteristics.
\begin{figure*}[ht]
\begin{picture}(18.4,14.0)
\put(0.0,0.0){\includegraphics[scale=1.0, angle=0, width=19cm]{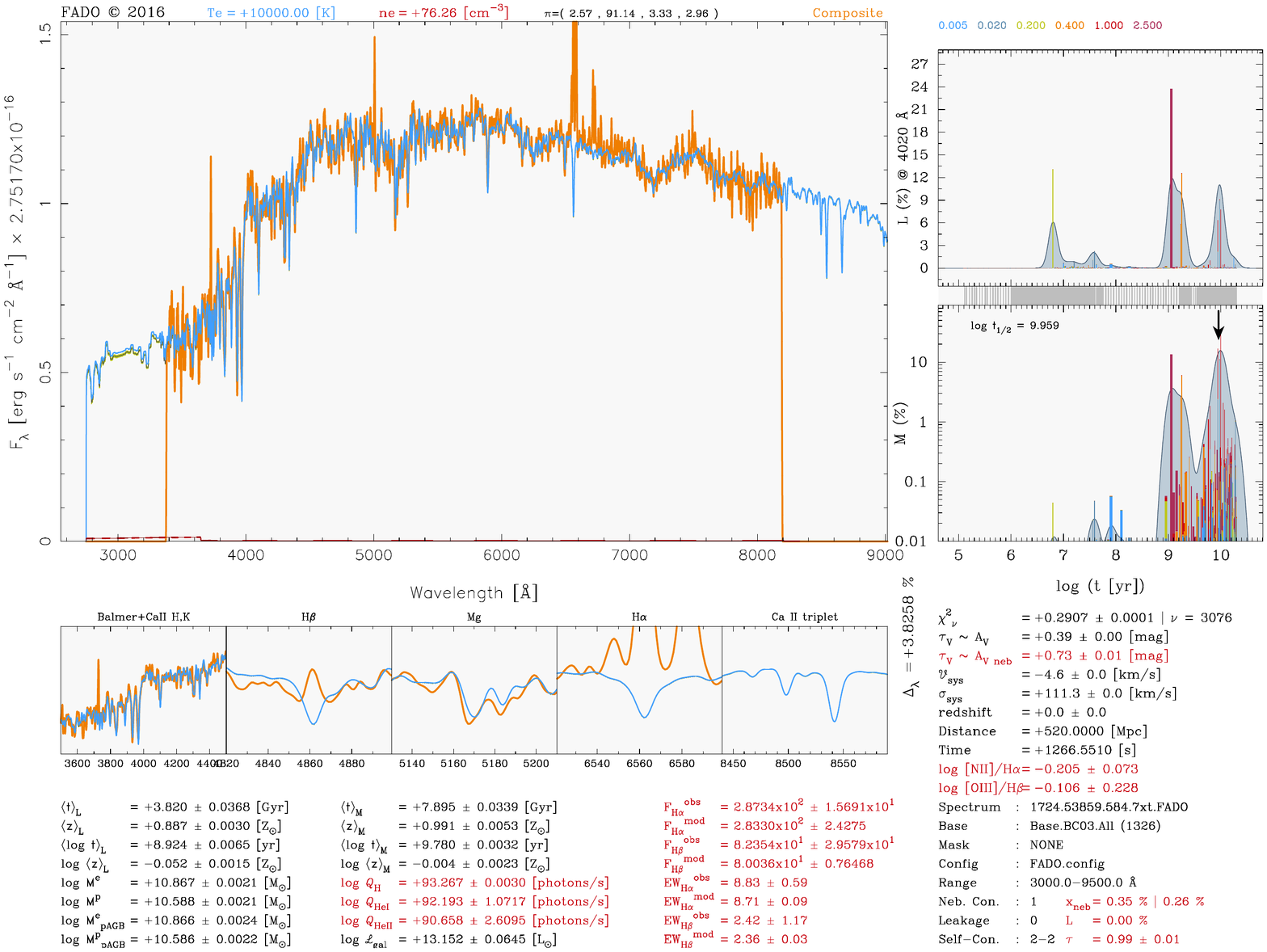}}
\PutLabel{1.6}{13.1}{\color{black}\huge a)}
\PutLabel{14.1}{12.7}{\color{black}\huge c)}
\PutLabel{1.6}{4.5}{\color{black}\huge b)}
\PutLabel{14.1}{6.7}{\color{black}\huge d)}
\end{picture}
\caption[]{Spectral model with \FADO\ of the "composite" galaxy
  (cf. Fig.~\ref{fig:BPT_WHAN}) \object{2MASX J15452094+0812244}. The meaning
  of the diagrams is identical to that in Fig.~\ref{fig:fit_SF} except for
  panel {\rem b}, which facilitates visual inspection of the quality of
  kinematical fitting of stellar absorption features within five selected
  spectral intervals. Additionally, we exclude from panels c\&d the vertical
  bars estimated $\pm$1$\sigma$ uncertainties for the luminosity and mass
  contribution of individual SSPs, in order to facilitate visual
  inspection. It can be seen that emission lines in this system are weak
  (EW(\ha)$\leq$9 \AA) and the nebular continuum is almost negligible
  throughout the considered spectral range.}
\label{fig:fit_Comp}
\end{figure*}
Last but not least, an important feature of \FADO\ is that, despite a severely
gas-contaminated spectral continuum, as in \object{CGCG 007-025}, it converges
into a sensible stellar velocity dispersion $\sigma$ ($\approx$90 km/s) being
within the range of values determined for dwarf galaxies
\citep[e.g.,][]{Guzman1998} . This is not the case for conventional
\pss\ codes where strong dilution of stellar absorption features by the
nebular continuum typically precludes computation of $\sigma$. This in turn
may either prevent smooth termination of the fitting procedure or drive
$\sigma$ to a maximum allowable bound, which could introduce a further bias in
the inferred SFH.  The so far poorly explored cumulative impact of the
aforementioned limitations and biases stemming from the neglect of nebular
continuum emission in currently available \pss\ codes is certainly a subject
of considerable relevance to studies of high-sSFR galaxies near and far (e.g.,
BCDs and green peas).
\begin{figure*}[ht]
\begin{picture}(18.4,14.0)
\put(0.0,0.0){\includegraphics[scale=1.0, angle=0, width=19cm]{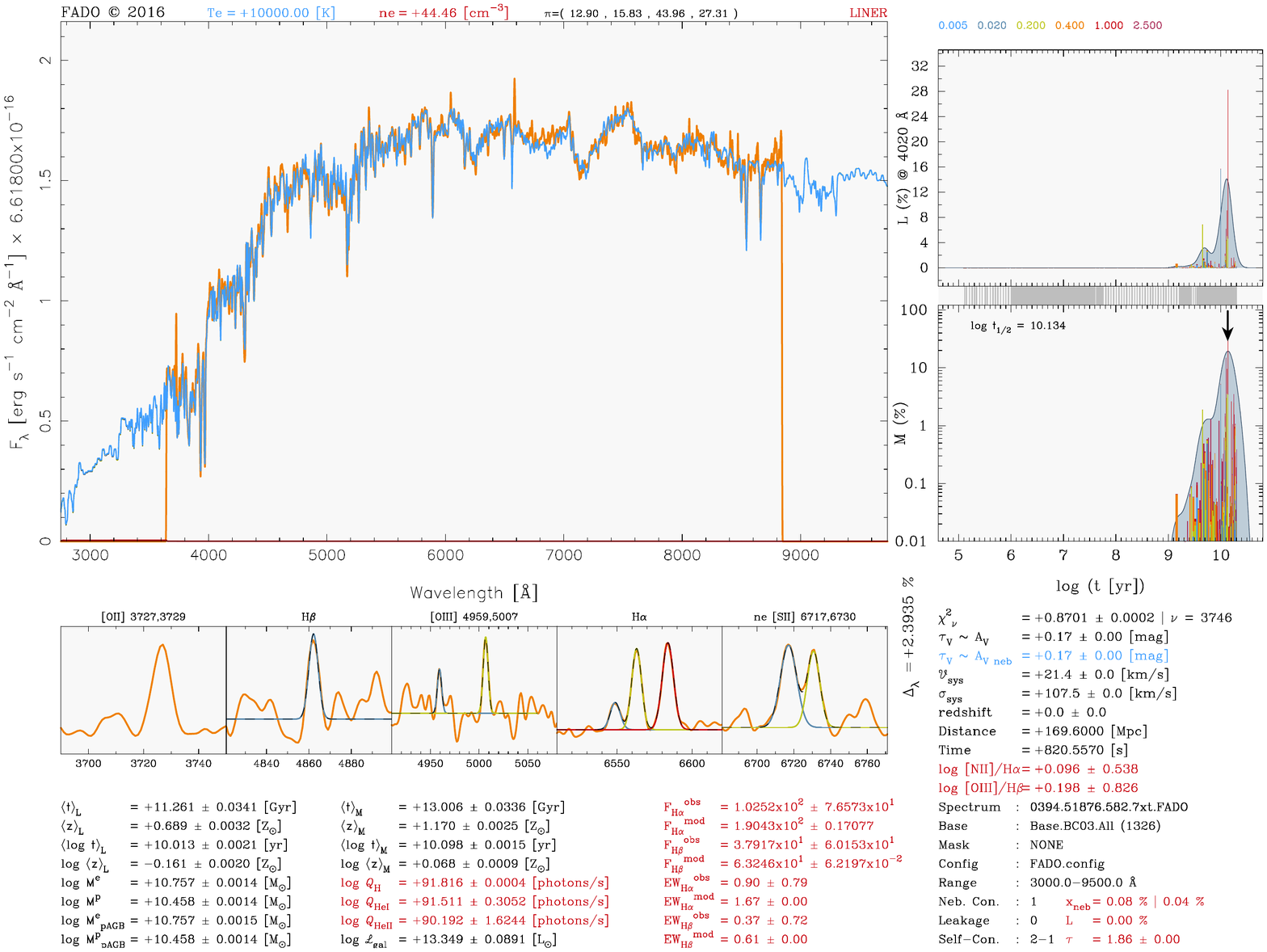}}
\PutLabel{1.6}{13.1}{\color{black}\huge a)}
\PutLabel{14.1}{12.7}{\color{black}\huge c)}
\PutLabel{1.6}{4.5}{\color{black}\huge b)}
\PutLabel{14.1}{6.7}{\color{black}\huge d)}
\end{picture}
\caption[]{Spectral model with \FADO\ of \object{MCG +00-03-030}, an
  early-type galaxy that falls in the locus of LINERs (passive/retired) in the
  BPT (WHAN) diagram (Fig.~\ref{fig:BPT_WHAN}). The layout is identical to
  that in Fig.~\ref{fig:fit_Comp} except for that in panel {\rem b}, which
  here displays magnified versions of Gaussian line fitting and deblending of
  weak emission lines with an EW$<$1 \AA.}
\label{fig:fit_ETG}
\end{figure*}

We turn next to an example of fitting with \FADO\ of a spectrum classified as
composite (Fig.~\ref{fig:fit_Comp}). It can be seen that emission lines in
this galaxy are weak, with the EW of the Balmer \ha\ line being lower than 9
\AA, and the EW(\hb) comparable to the value expected for the underlying
stellar absorption ($\sim$2 \AA). Despite the weakness of the \ha\ and
\hb\ lines, \FADO\ could maintain the \FCmode\ to full convergence (as
apparent from the agreement between observed and predicted emission-line
fluxes and EWs; third column from the left) and also indicated by the {\rem
  Self-Con} label marking the initial and final fitting mode (2 in either
case). However, since the auroral [O{\sc iii}]$_{4363}$ emission line could
not be detected in the observed spectrum, the electron temperature has been
fixed to the standard value (T$_{\rm e}=10^4$ K).

In Fig. \ref{fig:fit_ETG}, we show a spectral fit with \FADO\ of the
early-type galaxy (ETG) \object{MCG +00-03-030} that is classified
(probability 0.44) as LINER, as is typically the case for such systems
\citep[see e.g.,][and references therein]{Gomes2016}. This fit also uses an
extended library of 1326 SSPs from \citet{BruzualCharlot2003}.  As apparent
from comparison of panels {\rem a} and {\rem b}, emission lines in this galaxy
are very weak (0.9 \AA\ and 0.4 \AA\ for the EW(\ha) and EW(\hb),
respectively) and visible only after subtraction of the best-fitting SED from
the observed spectrum.  The faintness of emission lines obviously precludes
any accurate determination of T$_{\rm e}$ and n$_{\rm e}$, which are therefore
fixed to standard values ($10^4$ K and 100 cm$^{-3}$, respectively).  For this
reason, and because of the LINER nature of the source, \FADO\ switches from
the initially attempted \FCmode\ to the \NCmode, with other words it
drops the requirement for matching observed and predicted Balmer-line
luminosities and EWs, while maintaining the contribution of nebular continuum
in the SED fit.
 
Panel {\rem b} is chosen to show for the sake of visual quality control a
zoom-in into emission lines and their Gaussian fits. It can be appreciated
that, even in the concrete case of a low-S/N, weak-line spectrum, the genetic
DEO module of \FADO\ performs reasonably well in identifying and measuring
emission lines. Given, however, the large uncertainties in the \ha\ and
\hb\ fluxes, the extinction in the nebular component is fixed to the value
inferred for the stellar component. Note the strong discrepancy by a factor of
$\tau \sim 60$ between predicted and observed Balmer-line fluxes, which, taken
at face value, suggests that the bulk of the ionizing LyC radiation produced
by the stellar component escapes without being locally reproduced into nebular
emission \citep[][]{P13,Gomes2016}.

\begin{figure*}[ht]
\begin{picture}(18.4,14.0)
\put(0.0,0.0){\includegraphics[scale=1.0, angle=0, width=19cm]{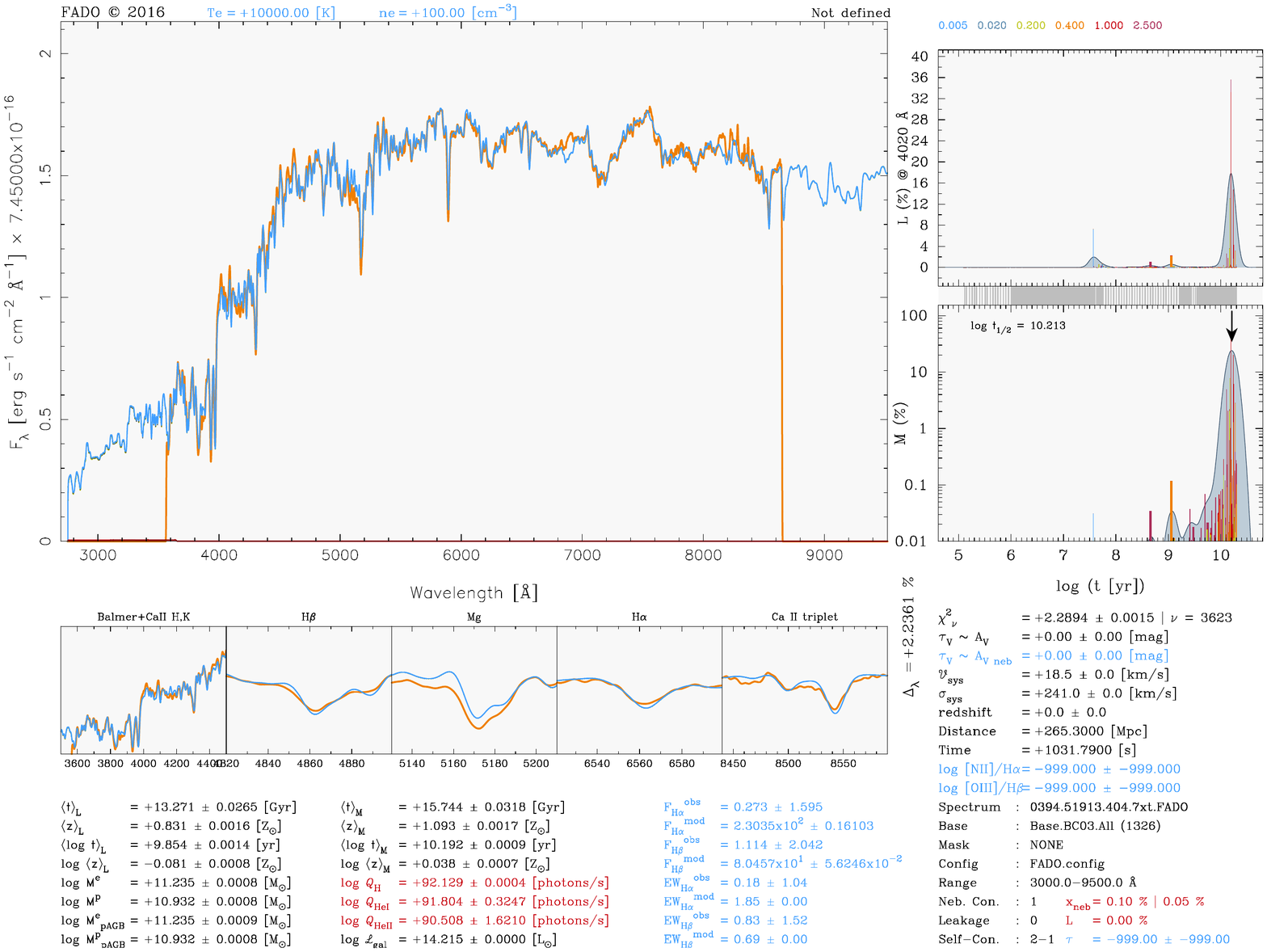}}
\PutLabel{1.6}{13.1}{\color{black}\huge a)}
\PutLabel{14.1}{12.7}{\color{black}\huge c)}
\PutLabel{1.6}{4.5}{\color{black}\huge b)}
\PutLabel{14.1}{6.7}{\color{black}\huge d)}
\end{picture}
\caption[]{Spectral model with \FADO\ of the lineless "passive" galaxy
  (cf. Fig.~\ref{fig:BPT_WHAN}) \object{III Zw 014}. The layout is as in
  Fig.~\ref{fig:fit_Comp}.}
\label{fig:fit_Passive}
\end{figure*}

Finally, Fig.~\ref{fig:fit_Passive} shows a model to the lineless
\citep["passive", in the notation by][]{Stasinska2008} galaxy \object{III Zw
  014}. It can be seen that the quality of kinematical absorption-line fitting
is satisfactory, with the exception of the Mg-feature, which reflects
well-known imperfections of currently available SSPs in reproducing
$\alpha$-enhancement features in ETGs \citep[e.g.,][see also
  \textcolor{blue}{Walcher et al. 2011} for a
  review]{WortheyFaberGonzalez1992,Worthey1992,Worthey1994}.

\section{Summary and conclusions \label{Conclusions} \label{Summary}}
In this article, we outline the physical and mathematical concept, and present
illustrative applications of \FADO\ (Fitting Analysis using Differential
Evolution Optimization), a novel publicly available population spectral
synthesis (\pss) tool.

\FADO\ opens a promising avenue to the exploration of the assembly history of
galaxies thanks to several unique features that go beyond existing concepts in
spectral synthesis and which substantially alleviate degeneracies and
systematic biases in state-of-the-art \pss\ codes, all of which neglect
nebular continuum emission and lack a mechanism that ensures consistency
between the best-fitting star formation history (SFH) and the observed nebular
emission characteristics. The main innovations embodied in \FADO\ pertain to
both its astrophysical self-consistency concept and its mathematical
realization: \\

{\rem i. Consistency between the best-fitting stellar model and the observed
  nebular emission characteristics in a galaxy:} A key innovation of
\FADO\ over all currently available \pss\ codes is {\rem a)} the inclusion of
the nebular continuum emission in spectral fits.  Nebular emission provides an
important fraction (on the order of 1/3) of the total optical/IR continuum
emission in galaxies undergoing strong star-forming activity, its
consideration is therefore fundamental to a realistic and unbiased spectral
modeling of such systems near and far; {\rem b)} consistency between the
observed nebular characteristics (luminosities and equivalent widths-EWs of
hydrogen Balmer emission lines, shape of the continuum around the Balmer and
Paschen jump) with the star formation- and chemical enrichment history
inferred from the best-fitting stellar model.

{\rem ii. Multi-objective optimization to Pareto solutions with Differential
  Evolution Optimization:} \FADO\ is the first \pss\ code in astrophysics
using an advanced variant of genetic Differential Evolution Optimization (DEO)
algorithms. This and various other elements in its mathematical foundation and
numerical realization (e.g., optimization of the spectral library used, test
for convergence through a procedure inspired by Markov Chain Monte Carlo
techniques) ensure quick convergence to the Pareto optimal solution. This,
together with the stability, high computational efficiency (internal
quasi-parallelization) and modular architecture of \FADO\ facilitates its
non-supervized application to large spectroscopic data sets and eases
future upgrades (e.g., integration of peripheral modules that permit a refined
treatment of nebular physics and infrared emission by dust).

Quite importantly, \FADO\ incorporates within a single code the entire chain
of data pre-processing, modeling, post-processing and graphical representation
of the results from \pss, including their storage in a convenient (FITS and
postscript) format. This integrated concept greatly simplifies and accelerates
a lengthy sequence of individual time-consuming steps generally involved in
\pss modeling, this way further enhancing the overall efficiency of \FADO.
Starting from the pre-processing of an input spectrum (flagging of emission
lines and spurious spectral features and, optionally, flux-conserving
rebinning, redshift determination and estimation of the error spectrum),
\FADO\ uses Artificial Intelligence concepts for an initial spectroscopic
classification and optimization of the library of Simple Stellar Population
(SSP) spectra subsequently used for spectral fitting.  Another unique feature
of \FADO\ is the estimation of uncertainties for all primary quantities that
define the best-fitting {\sl population vector}, that is the mass and
light contributions of individual SSPs involved in the fit, the intrinsic
extinction (both for the stellar and nebular component) and the velocity
dispersion. Uncertainties are computed and exported also for all secondary
products from the model. These include {\rem a)} various physical and
evolutionary characteristics of a galaxy spectrum (e.g., luminosity- and
mass-weighted stellar age and metallicity, ever formed and presently available
stellar mass and the expected hydrogen- and helium-ionizing Lyman continuum
photon rate from it, the time when a galaxy has assembled 1/2 of its
present-day stellar mass, and others) and {\rem b)} Fluxes and EWs, even for
faint (EW$\la$1~\AA) emission lines in the optical spectral range.  Besides
storage of the relevant primary and secondary output from spectral modeling
(into FITS format) for subsequent analysis, \FADO\ also facilitates graphical
visualization of its output (e.g., residuals between input spectrum and model,
kinematics of the stellar component, emission-line fitting and
deblending) through a built-in plotting routine.

Furthermore, \FADO\ incorporates already in its current publicly available
version a number of additional features (currently under testing and
fine-tuning) which will be offered in subsequent releases. These include the
allowance for the estimation of the Lyman continuum photon escape fraction and
the provision for a user-supplied instrumental line spread function for the
sake of improved kinematical fitting and decomposition.  Additionally, the
presently unique ability of \FADO\ to handle spectral libraries with up to
2000 SSPs with a dimension of up to 24k each opens opportunities for the
exploration with future higher-resolution SSPs of various topical questions in
extragalactic research, such as, e.g., a possible non-universality of the
stellar initial mass function or the $\alpha$-element enhancement in
early-type galaxies.
 
\newpage
\begin{acknowledgements}
We would like to thank the anonymous referee for numerous valuable comments
and suggestions.  This work was supported by Funda\c{c}\~{a}o para a
Ci\^{e}ncia e a Tecnologia (FCT) through national funds and by FEDER through
COMPETE by the grants UID/FIS/04434/2013 \& POCI-01-0145-FEDER-007672 and
PTDC/FIS-AST/3214/2012 \& FCOMP-01-0124-FEDER-029170. We acknowledge support
by European Community Programme ([FP7/2007-2013]) under grant agreement
No. PIRSES-GA-2013-612701 (SELGIFS).  J.M.G. was supported by the fellowship
SFRH/BPD/66958/2009 funded by FCT (Portugal) and POPH/FSE (EC) and by the
fellowship CIAAUP-04/2016-BPD in the context of the FCT project
UID/FIS/04434/2013 \& POCI-01-0145-FEDER-007672.  P.P. was supported by FCT
through Investigador FCT contract IF/01220/2013/CP1191/CT0002.  We thank
Mayanna Gomes for the invaluable discussions related to the field of genetics
and Leandro Cardoso for extensive tests of \FADO.
The graphical output (encapsulated postscript) from FADO (v.1) is produced
with PGplot (http://www.astro.caltech.edu/$\sim$tjp/pgplot), for which we
would like to thank Prof. T.J. Pearson and several people who have contributed
to the development of this graphics package.  This research has made use of
the NASA/IPAC Extragalactic Database (NED) which is operated by the Jet
Propulsion Laboratory, California Institute of Technology, under contract with
the National Aeronautics and Space Administration.
\end{acknowledgements}

\twocolumn



\appendix
\section{Illustrative comparisons of \FADO\ with state-of-the-art (purely stellar) population spectral synthesis codes \label{appendix}}

In this section we provide an illustrative comparison of results from
\FADO\ and \starlight\ \citep[][public distribution v04]{CidFernandes2005}.
The latter may be considered as representative of state-of-the-art
\pss\ codes, which without exception fit galaxy SEDs with purely stellar
templates and lack a mechanism that ensures consistency between the observed
nebular emission characteristics with the star-formation and chemical
enrichment history encoded in best-fitting PVs (Sect.~\ref{intro}).

\begin{figure*}[ht]
\begin{picture}(18.4,14.0)
\put(0.0,0.0){\includegraphics[scale=1.0, angle=0, width=19cm]{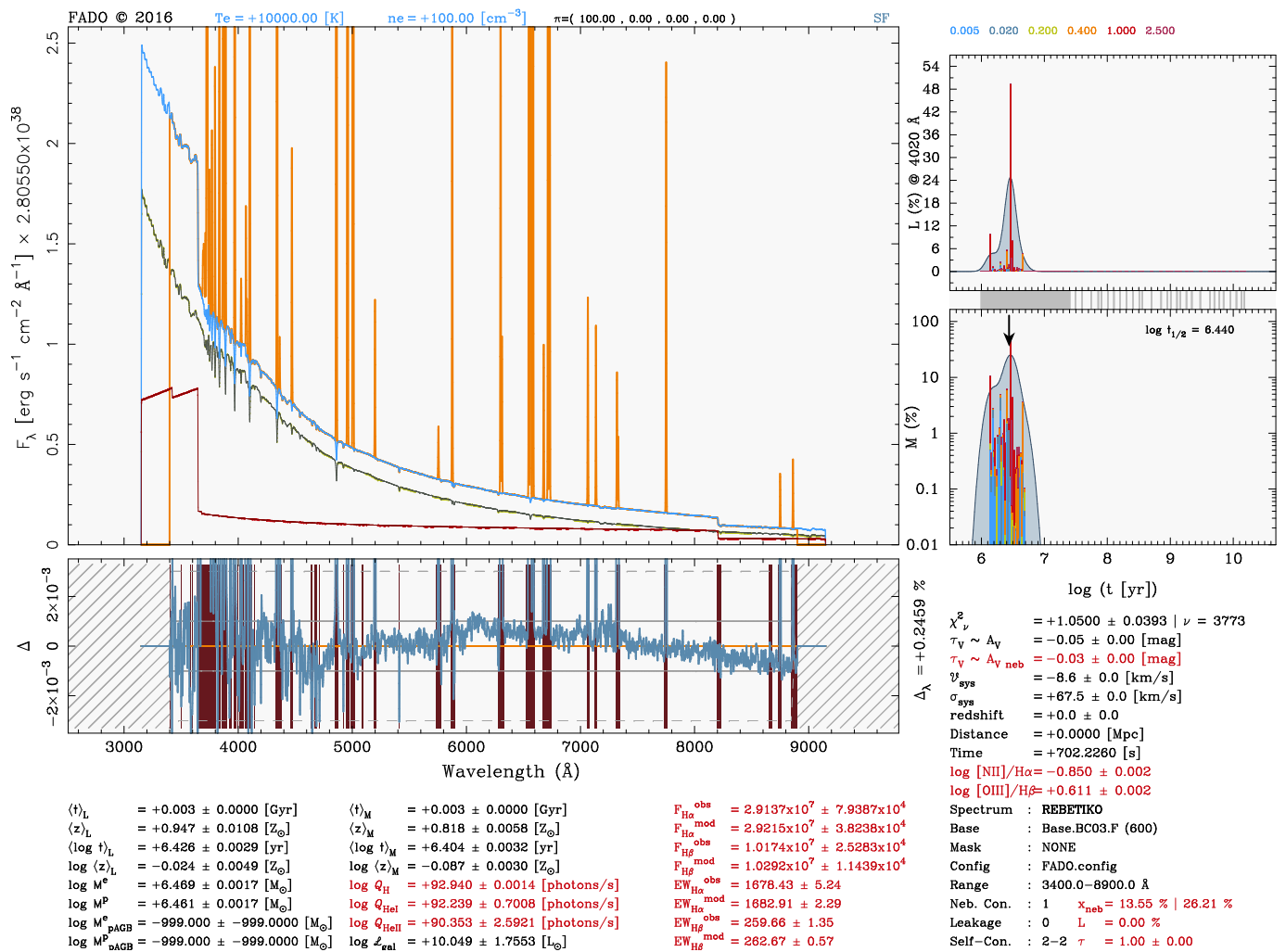}}
\end{picture}
\caption[]{Spectral fit with \FADO\ to a synthetic spectrum from {\sc
    Rebetiko} corresponding to an instantaneous-burst age of 3 Myr. The input
  spectrum includes nebular line and continuum emission computed from the LyC
  photon output from the stellar component, whereby case~B recombination and
  standard conditions for the gas were assumed. The fit was computed in the
  \FCmode\ of \FADO\ with the SSP library \BaseF.  The meaning of the panels
  is identical to that in Fig.~\ref{fig:fit_SF}.}
\label{RebetikoSpectrum}
\end{figure*}

Both \pss\ codes were applied to synthetic spectra computed with our
evolutionary synthesis code {\sc Rebetiko} (Papaderos \& Gomes, in prep.) for
two SFHs, one involving continuous star formation at a constant SFR and the
other a single instantaneous burst that is approximated by an exponentially
decaying SFR $\propto \rm{exp}(-t/\tau)$ with a short e-folding
timescale of $\tau=1$ Myr.  For each SFH synthetic spectra were computed
in the spectral range between 2500 -- 10000 \AA\ for 716 ages between 1~Myr
and 15~Gyr, assuming solar metallicity ($Z_{\odot}=0.02$) and zero intrinsic
extinction ($A_V=0$ mag).  The SSP templates used by {\sc Rebetiko} are from
\citet{BruzualCharlot2003}, and assume a \citet{Chabrier2003} IMF between 0.1
and 100 $M_{\odot}$ and Padova evolutionary tracks
(\citealt{Alongi1993,Bressan1993,Fagotto1994a,Fagotto1994b,Fagotto1994c,Girardi1996}).
For each synthetic SED, the LyC photon output from the stellar component
\citep[][]{Zanstra1961} was converted into nebular continuum assuming case~B
recombination and standard conditions (T$_{\rm e}$=10$^4$ K, n$_{\rm e}$=100
cm$^{-3}$; cf. Sect.~\ref{FADO:EmLin}). Additionally, {\sc Rebetiko} exports
into the synthetic SED hydrogen line fluxes computed for different electron
temperatures and densities \citep{Hummer1987}, as well as lines for heavier
elements adopted from the semi-empirical calibration by \citet{AF03} as a
function of the nebular metallicity.  The synthetic stellar + nebular spectra
from {\sc Rebetiko} were finally smeared to an intrinsic velocity dispersion
of 50 km/s both in the stellar and nebular component, and sightly degraded to
a S/N=200 at $\lambda_0=4020$ \AA.

In a second stage, the synthetic SEDs were modeled with \starlight\ and
\FADO\ using a base that comprises SSPs from \cite{BruzualCharlot2003} for 25
ages between 1 Myr and 18 Gyr and six metallicities ($Z = 0.005$, 0.02, 0.2,
0.4, 1 and 2.5 $Z_{\odot}$), hereafter \BaseS.  Fits were performed after
flagging of emission lines in the spectral range 3400 -- 8900 \AA\ following
common practice in studies of local galaxy samples from SDSS
\citep[e.g.,][]{Asari2007,Ribeiro2016} with the $A_V$ kept as a free parameter
between -1 and 4 mag.  We note that, since \FADO\ allows a maximum of
2000 SSPs in the base library (Sect.~\ref{FADO_in_a_nutshell}), whereas the
base in the case of \Starlight\ is limited to 300 elements, a comparison
between the two codes is only possible for SSP bases restricted to $\leq$300
elements.

Given that the computation of model uncertainties is not fully implemented in
\starlight, each spectrum was fitted ten times with different initial guesses
in the parameter space (seed numbers in the fitting procedure) to evaluate
formal errors \citep[see, e.g.,][]{Ribeiro2016,Cardoso2016-AGNL}.
Spectral fits with \FADO\ were computed in the \FCmode\ and with the default
number of evolutionary threads ({\bf 7}; cf. Sect.~\ref{FADO:DEO}).

In the following, we restrict the comparison between the two \pss\ codes to
only a few physical quantities that one generally seeks to determine through
\pss\ modeling of galaxy spectra: the stellar mass \mstar, light- and
mass-weighted mean stellar age (\tlum\ and \tmass, respectively), and light-
and mass-weighted mean stellar metallicity (\Zlum\ and \Zmass, respectively).
Additionally, we compare the \ha\ equivalent width of the input synthetic
spectra with that implied by the best-fitting PV from \FADO\ as an additional
means for evaluating the ability of the code to reproduce the nebular
characteristics of a galaxy spectrum in its full-consistency fitting mode
(\FCmode). The reader is referred to CGP17 for detailed CPU-time benchmark
tests of \FADO\ and a comparison of its output to that from currently
available \pss\ codes for synthetic galaxy spectra covering a wide range of
SFHs, evolutionary stages and levels of nebular emission contribution.  CGP17
also provide comparisons between the three fitting modes of
\FADO\ (Sect.~\ref{FADO:EmLin}) for synthetic spectra of various
characteristics (S/N ratio, wavelength coverage, spectral classification) and
different model setups (e.g., number of evolutionary threads, dimension of the
SSP library).

Figure \ref{RebetikoSpectrum} shows a fit with \FADO\ to a synthetic spectrum
from {\sc Rebetiko} for an instantaneous burst model of age 3 Myr. The figure
layout is identical to that in Fig.~\ref{fig:fit_SF}. It can be seen that the
best-fitting PV reproduces the Balmer and Paschen discontinuity (3646 and 8207
\AA, respectively), and implies \ha\ and \hb\ EWs (1682.9 and 262.7 \AA,
respectively) that closely match those of the input spectrum (1678.43 and
259.66 \AA, respectively). Quite importantly, despite the strong nebular
continuum contamination ($\sim$15 \% of the monochromatic luminosity at
$\sim$4020 \AA) the mass- and luminosity-weighted age and metallicity of the
stellar component are recovered by the model to within 5\% difference.

\begin{figure*}[ht]
\begin{picture}(10.0,14.0)
\put(0.0,0.0){\includegraphics[trim={0cm 9.0cm 0cm 0cm},clip,scale=1.0,angle=0,width=\textwidth]{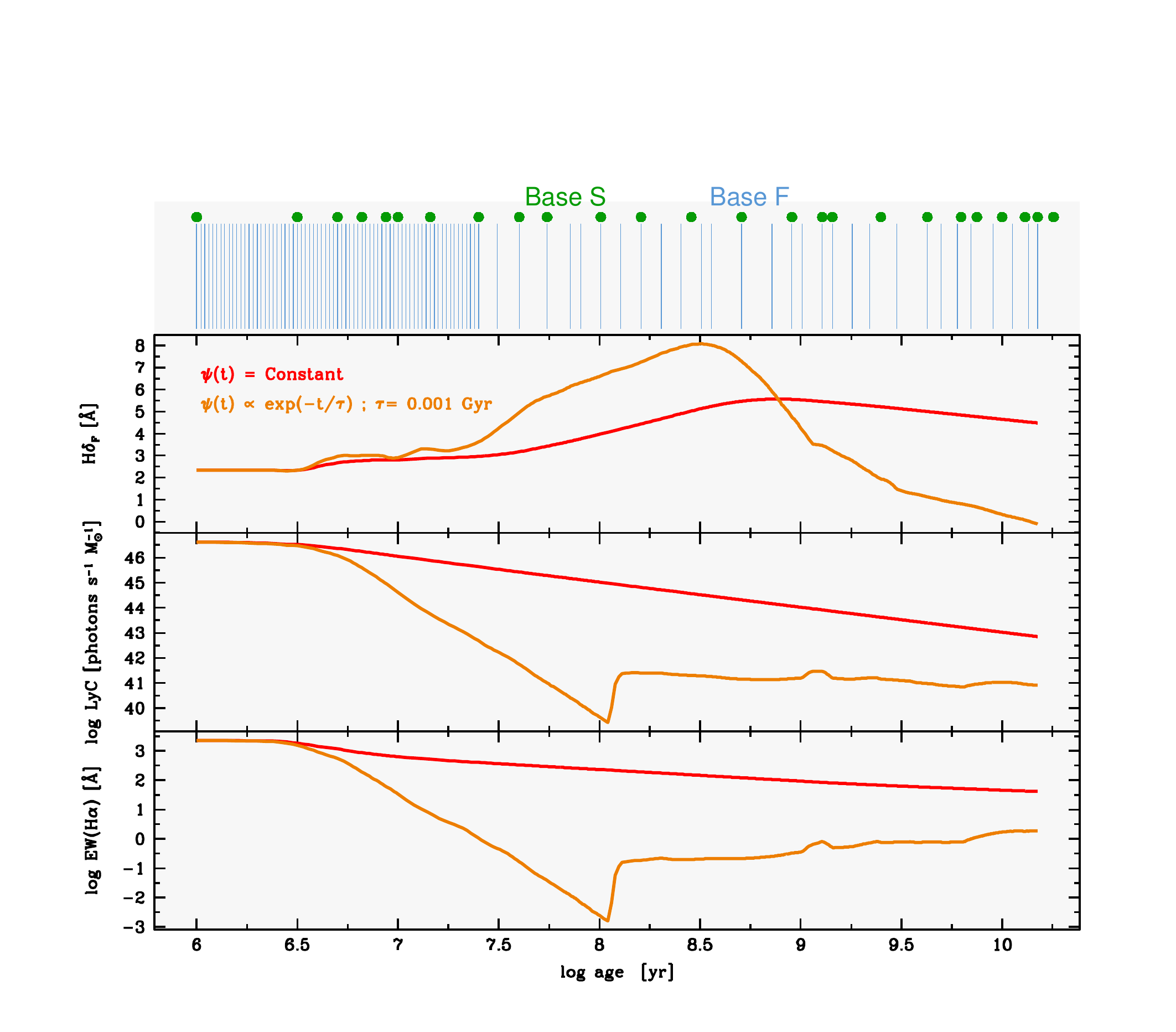}}
\put(16.0,10.2){\color{black}\huge a)}
\put(16.0,06.9){\color{black}\huge b)}
\put(16.0,03.9){\color{black}\huge c)}
\end{picture}
\caption[]{Time evolution computed with {\sc Rebetiko} of {\bf a)} the Lick
  index \hd$_F$, {\bf b)} LyC photon rate per unit mass and {\bf c)}
  \ha\ emission-line equivalent width for an instantaneous burst (orange) and
  continuous star formation at constant SFR (red).  The vertical blue strips
  on top of the panels mark the SSP ages for five stellar metallicities
  included in the \FADO-optimized library \BaseF, which was compiled such as
  to simultaneously yield an adequate coverage of steep decline in the LyC
  photon rate in the first $\sim$40 Myr and the significant evolution of EWs
  of stellar absorption features (e.g., the \hd$_F$ Lick index) over the
  ensuing $\sim$1 Gyr in the evolution of an instantaneously formed stellar
  population.  The SSP ages in \BaseS\ (green filled circles) are included for
  comparison.}
\label{A:Hdelta}
\end{figure*}

Despite a long-standing controversy on inherent algebraic degeneracies in
\pss\ \citep[e.g.,][]{Worthey1992,OConnell1996,Pelat1997,Pelat1998}, a subtle
yet important aspect in the periphery of this debate pertains to the
construction of the SSP base.  As we will argue below (and in more detail in
CGP17), the commonly used small-to-medium size bases that generally comprise a
limited number ($\la200$ elements) of hand-picked SSPs is one of the causes
for induced degeneracies in \pss. This is because such bases are typically
designed having in mind a reasonable trade-off between age/metallicity
coverage and CPU-time expense.  This task is in some cases aided by
mathematical tools intended to eliminate nearly redundant elements in the SSP
base and the grouping of base elements with similar properties into a single
representative block using diffuse K-means algorithms
\citet[e.g.,][]{Richards2009}.  However, this approach is highly
dependent on the number of cluster nodes\footnote{In general, the cluster
  centers in the supervized learning techniques are the averaged
  properties of spectra belonging to that particular node, that is the
  ensemble of spectra in a given node is represented by a mean SSP spectrum.}
spanning the base for spectral fitting.  More importantly, a caveat of this
approach is that depending on the number and selection criteria for cluster
centers, short evolutionary phases that could hold decisive indicators that
help constraining the fit (e.g., the steeply rising stellar H$\delta$
absorption-line EW in post-starburst galaxy spectra) might not be adequately
covered by the base, which could then impact the best-fitting PV in a
systematic or non-predictable manner, depending on the characteristics of the
studied spectrum and the design of the SSP base. One simple example for this
is that an attempt to fit a blue starburst-galaxy spectrum with a base lacking
young SSPs might force the \pss\ code to invoke a negative intrinsic
extinction, just like in the case when a strongly subsolar (thus bluish)
stellar component is modeled with solar/super-solar (thus reddish) SSPs. In
other cases, however, for instance, when fitting of a post-starburst spectrum
exhibiting strong ($\la$--6 \AA) H$\delta$ stellar absorption features using a
base that does not adequately cover ages between $\sim$0.1 and $\sim$0.7 Gyr,
it is difficult to predict the resulting biases in the best-fitting PV.
Another element to consider when aiming at consistency between the
best-fitting PV and the observed nebular characteristics of a galaxy, as in
the case of \FADO, is that instantaneously formed stellar populations
(i.e., SSPs) show during early evolutionary stages (1--40 Myr) a
relatively similar SED slope in the optical, but imply vastly different LyC
photon rates, and in consequence of this, different hydrogen Balmer-line
luminosities. A dense age (and metallicity) coverage for young SSPs is
therefore essential for realizing and exploiting the self-consistency concept
of \FADO\ in its \FCmode\ and \NCmode.  A mathematically and astrophysically
motivated construction of extended SSP bases that allows coverage in
time and metallicity of phases associated with a strong evolution in the LyC
photon production rate and the optical spectrophotometric characteristics of
stellar populations is therefore crucially important. At the same time, large
SSP libraries designed on the above criteria are well adapted to the
capability of \FADO\ to cope with bases of up to 2000 elements.

From such considerations, we have selected SSPs to produce a \FADO-optimized
base (\BaseF) that densely traces the rapid decrease of the LyC output and the
strong variation of the \hd$_F$ Lick absorption index in early ($\leq$40 Myr)
and intermediate (0.1-1 Gyr) evolutionary stages (see
Fig.~\ref{A:Hdelta}). Ages of SSPs in \BaseF\ are depicted in the top of
the figure, whereas filled green circles mark the ages composing the smaller
SSP library \BaseS.  \BaseF, containing 600 elements (100 ages between 1 Myr
and 15 Gyr for six metallicities, $Z = 0.005$, 0.02, 0.2, 0.4, 1 and 2.5
$Z_{\odot}$) was used for modeling with \FADO\ synthetic spectra computed with
{\sc Rebetiko} (e.g., Fig. \ref{RebetikoSpectrum}).

\begin{figure*}[!t]
\begin{picture}(18.0,15.0)
\put(0.0,8.0){\includegraphics[trim={0.2cm 7.05cm -0.75cm 13.0cm},clip,scale=1.0,angle=0, width=\textwidth]{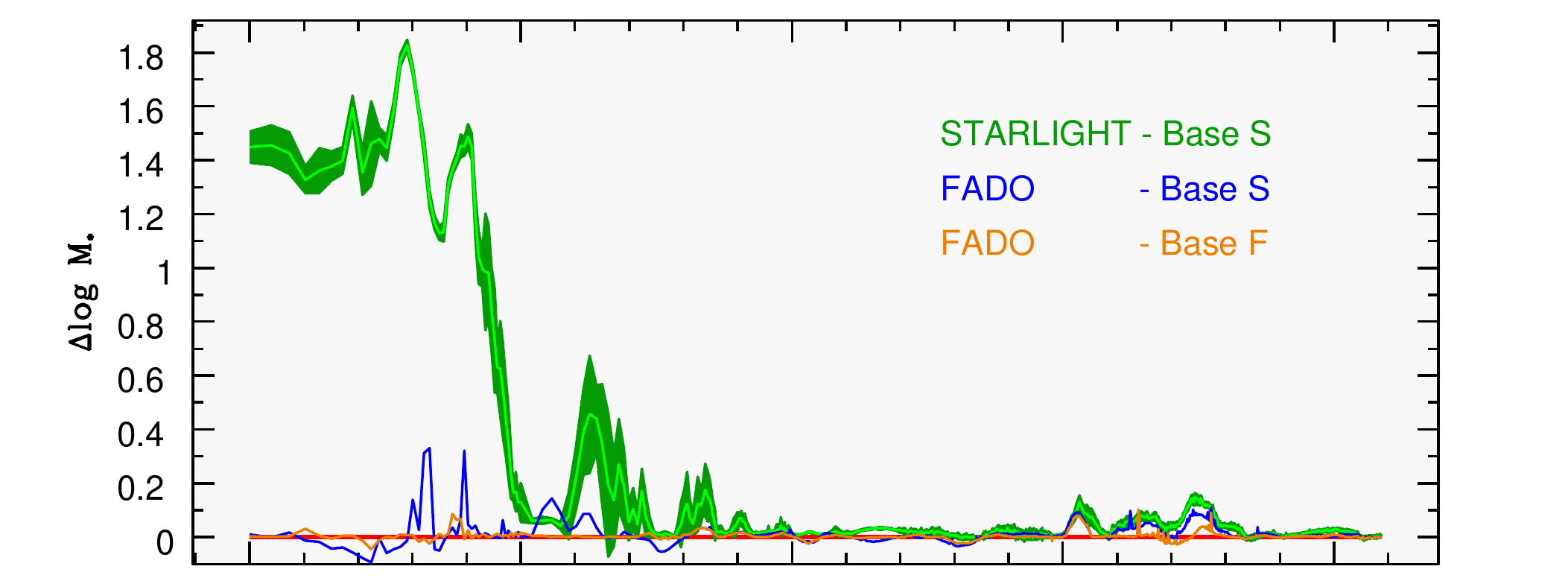}}
\put(0.0,0.0){\includegraphics[trim={0.2cm 5.55cm -0.75cm 13.0cm},clip,scale=1.0,angle=0, width=\textwidth]{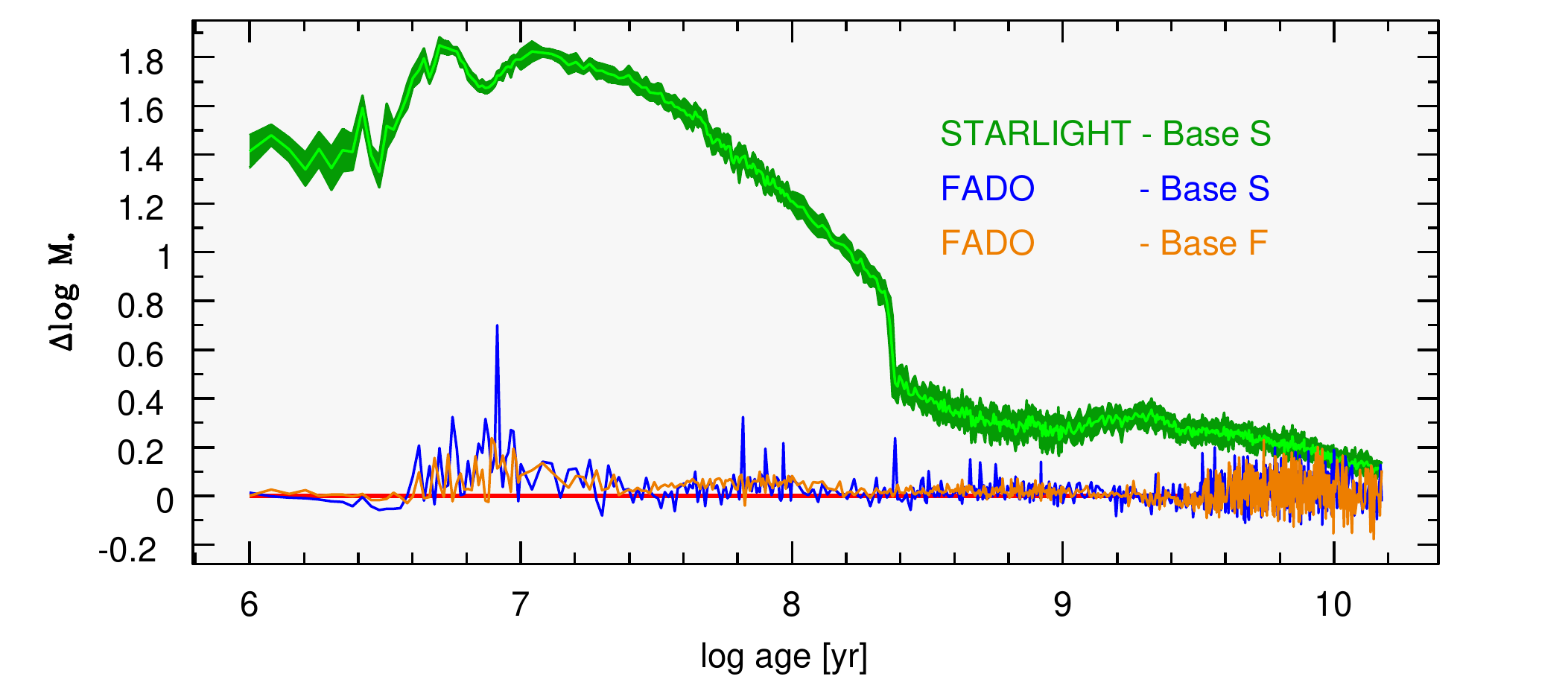}}
\PutLabel{15.5}{13.7}{\color{black}\huge a)}
\PutLabel{15.5}{6.70}{\color{black}\huge b)}
\end{picture}
\caption[]{Logarithmic representation of the difference between output and
  input stellar mass \mstar\ for spectral fits with \FADO\ and
  \starlight\ (blue and green, respectively) to synthetic spectra
  corresponding to {\rem{a)}} a single-burst star formation scenario
  approximated by a declining SFR given by exp$(-t/\tau)$ with an e-folding
  timescale $\tau = 1$ Myr and {\rem{b)}} continuous star formation at
  constant SFR. The synthetic input spectra assume solar-metallicity and zero
  intrinsic extinction, and include the contribution of nebular emission
  (continuum and lines).  The shaded green area around determinations with
  \starlight\ for the \BaseS\ SSP library depict $\pm$1$\sigma$ uncertainties,
  as obtained from ten fits.  A substantial contribution of the reddish
  nebular continuum to the SED for ages lower than $\sim 10^7$ yr in panel a
  and lower than $\sim 10^8$ yr in panel b causes the purely stellar fits by
  \starlight\ to strongly overestimate \mstar\ by up to $\sim$1.8 dex. To the
  contrary, \FADO\ can recover \mstar\ to within less than $\sim$0.5 dex over
  the entire age interval considered. Results for spectral fits with \BaseF
  are included in orange for the sake of completeness, but only for models
  with \FADO\ since this base contains 600 elements, twice the maximum number
  of base elements that \starlight\ can deal with.  Note the significant
  improvement of the results from \FADO\ when \BaseF\ is used instead of
  \BaseS, which is due to a better time coverage of evolutionary phases with a
  strong variation in the LyC photon rate and the \hd$_F$ absorption Lick
  index.}
\label{A:Mass}
\end{figure*}

\begin{figure*}[!t]
\begin{picture}(18.0,15.0)
\put(0.0,8.0){\includegraphics[trim={0.2cm 7.05cm -0.75cm 13.0cm},clip,scale=1.0,angle=0, width=1.00\textwidth]{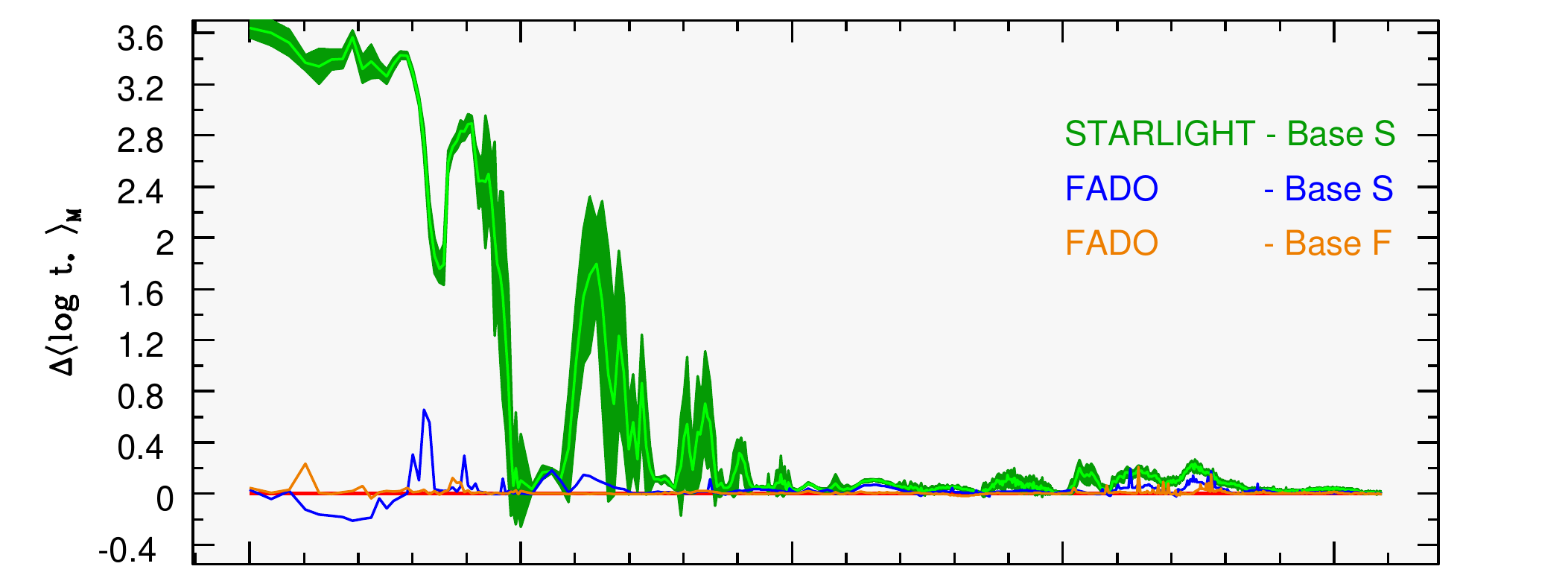}}
\put(0.0,0.0){\includegraphics[trim={0.2cm 5.55cm -0.75cm 13.0cm},clip,scale=1.0,angle=0, width=1.00\textwidth]{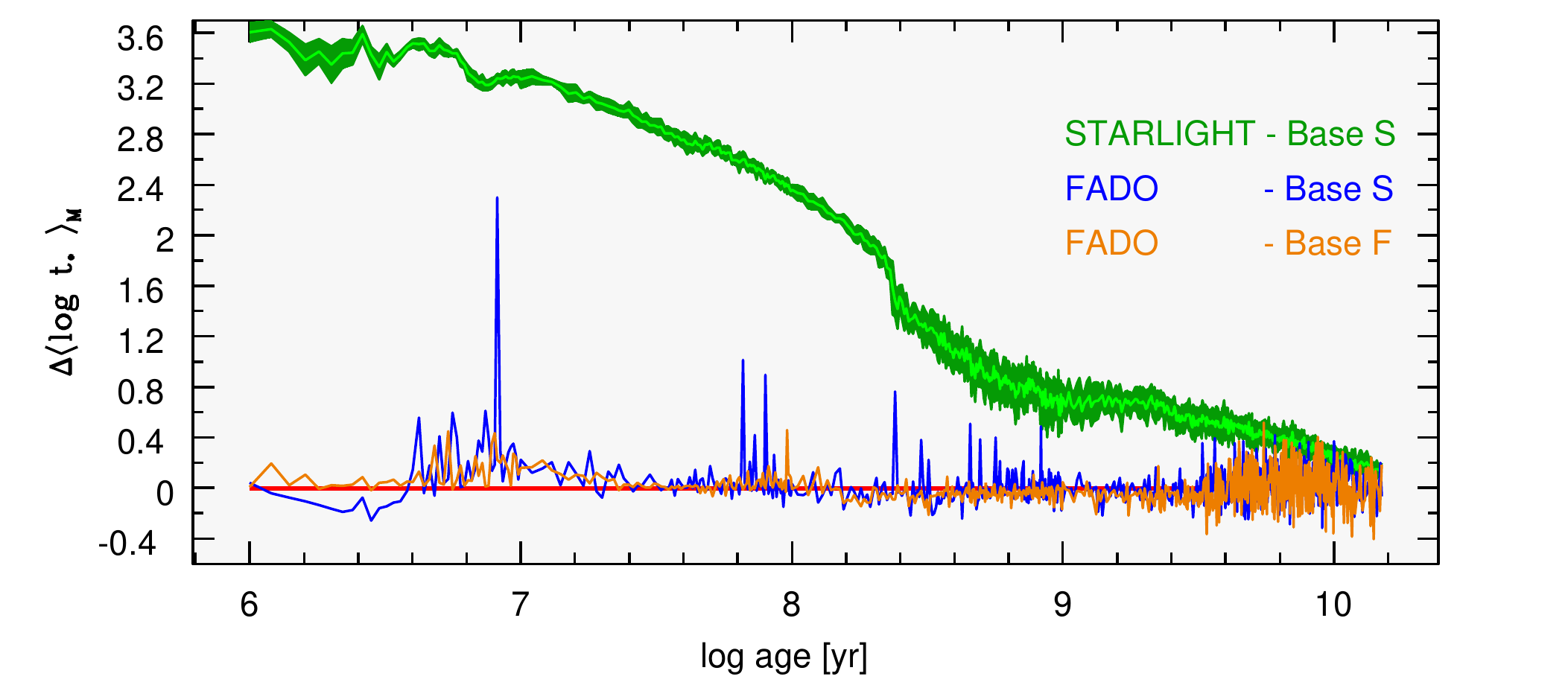}}
\PutLabel{15.5}{13.7}{\color{black}\huge a)}
\PutLabel{15.5}{7.00}{\color{black}\huge b)}
\end{picture}
\caption[]{Difference between output and input of the mean stellar age
  \tmass\ for synthetic spectra modeled with \FADO\ and \starlight. The colors
  have the same meaning as in Fig.~\ref{A:Mass}.}
\label{A:MeanAgeMass}
\end{figure*}

Figure~\ref{A:Mass} shows as a function of evolutionary time the difference
between the inferred and true ever formed \mstar\ for synthetic spectra
modeled with \FADO\ and \starlight\ (green and blue, respectively). Spectral
fits in either case were computed with the \BaseS\ library. The shaded green
area around determinations from \starlight\ depict $\pm$1$\sigma$
uncertainties estimated from ten model runs.  As a comparison, results from
\FADO\ for the \BaseF\ library are overlaid in orange.  The most salient
feature from this diagram is that purely stellar fits with
\starlight\ strongly overestimate \mstar\ by up to $\sim$1.8 dex for ages $\la
10^7$ and $\la 10^8$ yr for a single burst and constant SFR,
respectively. This bias is apparent over the period in which the nebular
contribution is significant (EW(\ha)$\gtrsim$100 \AA;
cf. Fig.~\ref{A:Hdelta}).

\begin{figure*}[t]
\begin{picture}(18.0,22.0)
\put(0,0){\includegraphics[trim={1cm 4cm -1cm -1cm},clip,scale=1.0,angle=0,width=20.0cm,height=25cm]{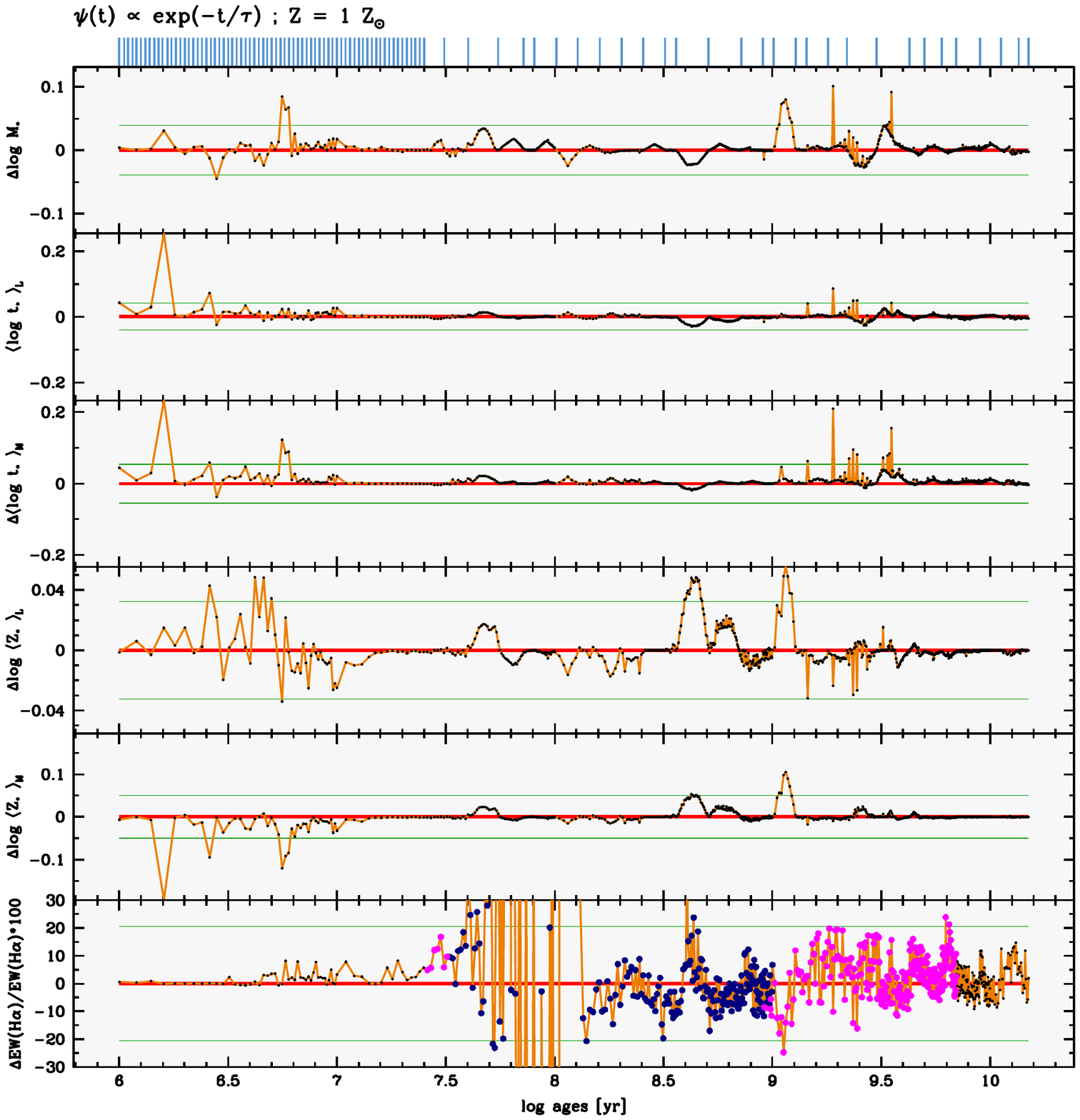}}
\PutLabel{16.8}{19.6}{\color{black}\huge a)}
\PutLabel{16.8}{16.7}{\color{black}\huge b)}
\PutLabel{16.8}{13.8}{\color{black}\huge c)}
\PutLabel{16.8}{11.0}{\color{black}\huge d)}
\PutLabel{16.8}{08.2}{\color{black}\huge e)}
\PutLabel{16.8}{05.3}{\color{black}\huge f)}
\end{picture}
\caption[]{Difference between output from \FADO\ for spectral fits with the
  \BaseF\ and input value of stellar mass \mstar\ (a), mean light- and
  mass-weighted stellar age (b\&c), mean light- and mass-weighted stellar
  metallicity (d\&e), and mean percentual deviation of the \ha\ emission-line
  equivalent width (f). The synthetic input spectra were computed with {\sc
    Rebetiko} assuming an instantaneous burst star formation scenario. In
  panel f the blue points correspond to equivalent-width of \ha\ in emission
  below 0.5\AA\ and the magenta points mark values in the range $0.5\AA \le$
  EW(\ha) $\le 1\AA$. The green lines are three times the standard
  deviation over the whole age range of the models. The vertical blue strips
  on top of the diagrams depict the ages of the SSPs composing \BaseF.}
\label{A:Burst}
\end{figure*}

\begin{figure*}[t]
\begin{picture}(18.0,22.0)
\put(0,0){\includegraphics[trim={1cm 4cm -1cm -1cm},clip,scale=1.0,angle=0,width=20.0cm,height=25cm]{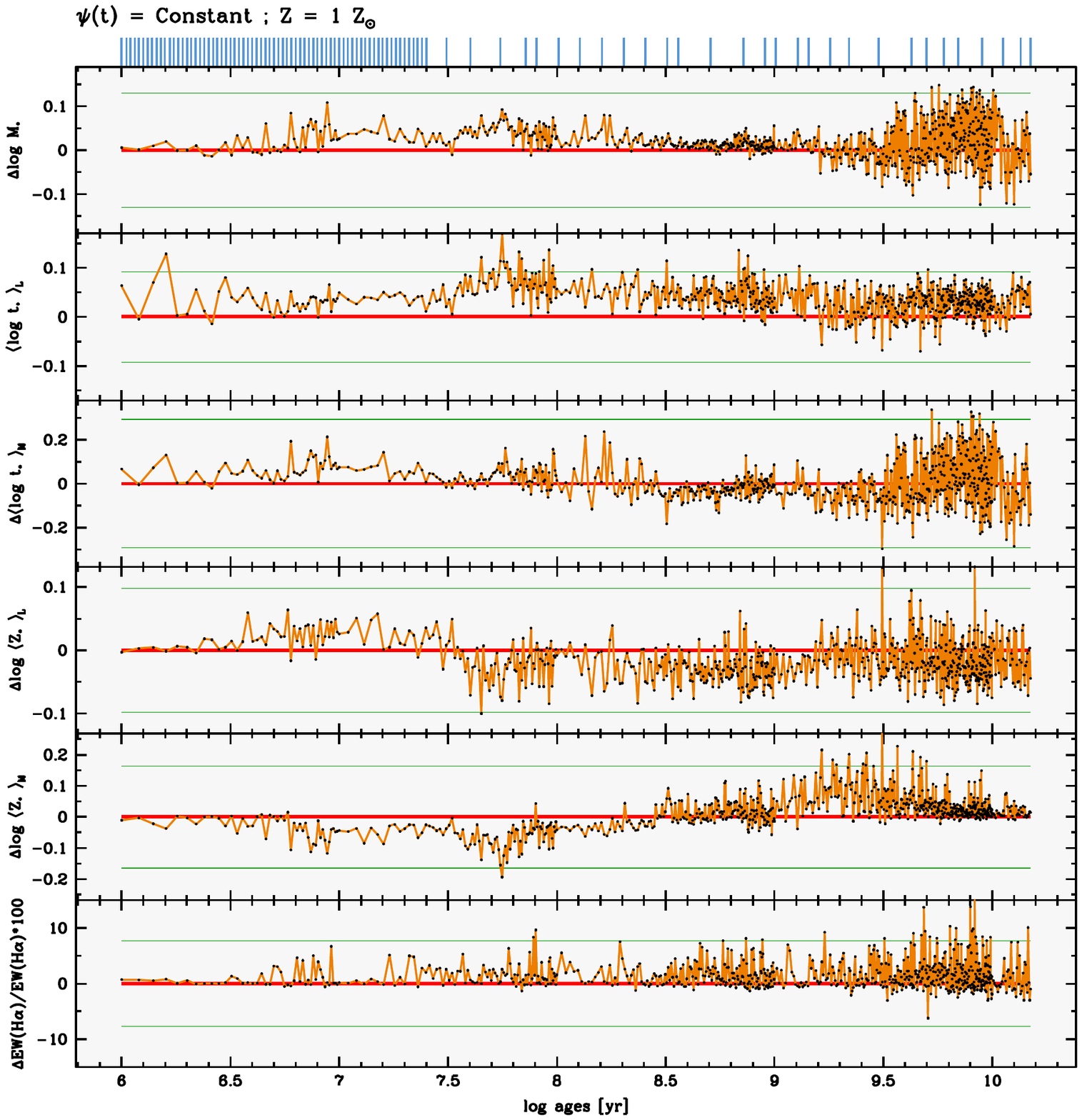}}
\PutLabel{16.8}{19.6}{\color{black}\huge a)}
\PutLabel{16.8}{16.7}{\color{black}\huge b)}
\PutLabel{16.8}{13.8}{\color{black}\huge c)}
\PutLabel{16.8}{11.0}{\color{black}\huge d)}
\PutLabel{16.8}{08.2}{\color{black}\huge e)}
\PutLabel{16.8}{05.3}{\color{black}\huge f)}
\end{picture}
\caption[]{As in Fig.~\ref{A:Burst}, but for synthetic spectra computed with
  {\sc Rebetiko} assuming continuous star formation at a constant SFR.}
\label{A:Continuous}
\end{figure*}

This fact is also echoed by Fig.~\ref{A:MeanAgeMass}, where we show as a
function of time the deviation of \tmass\ from the true value for
\starlight\ and \FADO\ fits.  \starlight\ strongly overestimates the mean
stellar age by up to 3.6 dex for ages $\la 10^7$ yr and $\la 10^8$ yr in the
case of an instantaneous burst and constant SFR. We would like to point
  out that, for the latter SFH, purely stellar fits with \starlight\ result
in a less prominent ($\la$0.8 dex) yet systematic overestimation of
\tmass\ for ages well beyond $> 10^9$ yr.  As for the results from \FADO,
singular peaks due to degeneracies induced by the construction of the
relatively small library \BaseS\ (see above) almost vanish when the
\FADO-optimized SSP library \BaseF\ is used. The insights from both
Fig.~\ref{A:Mass}\&\ref{A:MeanAgeMass} underscore the conclusion by
\citet[][see also \textcolor{blue}{Papaderos \& \"Ostlin 2012}]{Izotov2011-GP}
that neglect of the reddish nebular continuum emission leads spectral fitting
codes to infer a much too high fraction of old, high-M/L stars.

A more complete set of comparisons between \FADO\ (output) applied using the
\BaseF\ library to synthetic spectra from {\sc Rebetiko} (input) is shown in
Fig.~\ref{A:Burst} \&~\ref{A:Continuous}. These display for the same set of
spectral models the difference between output and true stellar mass ever
formed (a), light and mass-weighted stellar age (b\&c), light and
mass-weighted stellar metallicity (d\&e). These figures additionally include
the mean percentual deviation of the equivalent width of \ha\ (f), in order to
validate the full-consitency mode of \FADO. The uncertainties taking both set
of models into account and using \BaseF\ are on average less than 0.05 dex and
up to 0.15 dex for the stellar mass, on average less than 0.1 dex and up to
0.3 dex in the mean stellar age weighted both by light and
mass. Light-weighted stellar metallicities are recovered within $\sim$0.15 dex
and mass-weighted metallicities within $\sim$0.25 dex.  In its \FCmode,
\FADO\ allows one to self-consistently reproduce the Balmer
\ha\ equivalent width within a mean percentual deviation of 10\%. It is
  worth pointing out that the high mean percentual deviation of EW(\ha) in
Fig.~\ref{A:Burst} for ages $7.5 < $log(t)$ < 8.1$ is documented over an age
interval where EW(\ha) has dropped to $<$0.1 \AA, which means that it is
negligible in absolute terms.
 
\end{document}